\documentclass[letterpaper,twocolumn,10pt]{article}
\usepackage{usenix}
\usepackage[available]{usenixbadges}

\usepackage{filecontents}
\usepackage{cite}
\usepackage{indentfirst}
\usepackage{nicefrac}
\usepackage{gensymb}
\usepackage{siunitx}
\usepackage{bbding}
\usepackage{pifont}
\usepackage{pifont} 
\usepackage{array,framed}
\usepackage{booktabs}
\usepackage{seqsplit}
\usepackage{textgreek}
\usepackage{tabularx}
\usepackage{adjustbox}
\usepackage{minted} 
\usepackage{xcolor}
\definecolor{LightGray}{rgb}{0.97,0.97,0.97}
\usepackage{
  color,
  float,
  epsfig,
  wrapfig,
  graphics,
  graphicx
}
\usepackage{textcomp}
\usepackage{setspace}
\usepackage{amsmath}
\usepackage{amsfonts}
\usepackage[ruled,linesnumbered]{algorithm2e}
\usepackage{latexsym,fancyhdr}
\usepackage{enumerate}
\usepackage{algpseudocode}
\usepackage{xparse} 
\usepackage{xspace}
\usepackage{multirow}
\usepackage{tcolorbox}
\tcbuselibrary{breakable}
\usepackage{csvsimple}
\usepackage{balance}
\usepackage{hyperref}
\usepackage{cleveref}
\usepackage{
  tikz,
  pgfplots,
  pgfplotstable
}
\usepackage{pifont}

\usepackage{tikz}
\usepackage{cancel}
\usepackage{makecell}

\usepackage{caption}
\hypersetup{
    breaklinks=true,
    colorlinks=true,
    pdfborder={0 0 0}
}

\newif\ifshow
\showtrue

\DeclareRobustCommand{\responsebox}[2][gray!20]{%
\begin{tcolorbox}[
        breakable,
        left=0pt,
        right=0pt,
        top=0pt,
        bottom=0pt,
        colback=#1,
        colframe=#1,
        width=\linewidth, 
        enlarge left by=0mm,
        boxsep=2pt,
        arc=0pt,outer arc=0pt,
        ]
        #2
\end{tcolorbox}
}

\usepackage{listings}

\usetikzlibrary{
  shapes.geometric,
  arrows,
  external,
  pgfplots.groupplots,
  matrix
}
\pgfplotsset{compat=1.9}
\usepackage{mathtools}

\newcommand{\wb}[1]{\textcolor{black}{#1}}

\newcommand{\tool}{\textsc{PyGuard}}

\begin{document}

\title{Cutting the Gordian Knot: Detecting Malicious PyPI Packages via a Knowledge-Mining Framework}

\author{
{\rm Wenbo Guo}$^{1}$, 
{\rm Chengwei Liu}$^{2}$\thanks{Corresponding author}, 
{\rm Ming Kang}$^{3}$, 
{\rm Yiran Zhang}$^{1}$, 
{\rm Jiahui Wu}$^{1}$,\\
{\rm Zhengzi Xu}$^{4}$, 
{\rm Vinay Sachidananda}$^{1}$, 
{\rm Yang Liu}$^{1}$\\
$^{1}$Nanyang Technological University, 
$^{2}$Nankai University, \\
$^{3}$Sichuan University, $^{4}$Imperial Global Singapore
}

\maketitle

\begin{abstract}

The Python Package Index (PyPI) has become a target for malicious actors, yet existing detection tools generate false positive rates of 15-30\%, incorrectly flagging one-third of legitimate packages as malicious. This problem arises because current tools rely on simple syntactic rules rather than semantic understanding, failing to distinguish between identical API calls serving legitimate versus malicious purposes. To \wb{address} this challenge, we propose \tool{}, a knowledge-driven framework that converts detection failures into useful behavioral knowledge by extracting patterns from existing tools' false positives and negatives. Our method \wb{utilizes} hierarchical pattern mining to identify behavioral sequences that \wb{distinguish} malicious from benign code, employs Large Language Models to create semantic abstractions beyond syntactic variations, and combines this knowledge into a detection system that integrates exact pattern matching with contextual reasoning. \tool{} achieves 99.50\% accuracy with only 2 false positives versus 1,927-2,117 in existing tools, maintains 98.28\% accuracy on obfuscated code, and identified 219 previously unknown malicious packages in real-world deployment. The behavioral patterns show cross-ecosystem applicability with 98.07\% accuracy on NPM packages, demonstrating that semantic understanding enables knowledge transfer across programming languages.

\end{abstract}

\section{Introduction}
\label{sec:intro}

The Python Package Index (PyPI) serves as the central repository for Python packages, hosting over 400,000 packages with billions of downloads annually~\cite{pypi_stats_2024}. However, its open nature has made it an attractive target for malicious actors seeking to compromise software supply chains through typosquatting, dependency confusion, and code injection attacks~\cite{ohm2020backstabber, sejfia2020towards}. Recent studies document a substantial increase in malicious package uploads, with researchers identifying 116 malicious packages downloaded more than 10,000 times in 2023 alone~\cite{eset2023malicious}, including sophisticated attacks by groups like Lazarus that deployed typosquatting packages downloaded 300-1,200 times each~\cite{jpcert2024lazarus}, demonstrating how compromised packages can cascade through the entire software supply chain, affecting millions of downstream applications.

To combat malicious packages, detection approaches span three primary categories: static analysis tools (like Bandit~\cite{bandit2024} and GuardDog~\cite{guarddog2022}) that scan source code patterns with minimal overhead, dynamic analysis systems (like DySec~\cite{dysec2025}) that achieve 95.99\% accuracy by monitoring runtime behaviors, and machine learning methods (like PypiGuard~\cite{pypiguard2025}) that combine metadata with API behaviors to reach 98.43\% accuracy. However, these tools suffer from prohibitively high false positive rates, with current PyPI detection systems flagging approximately one-third of legitimate packages as malicious and generating false positive rates of 15-30\%~\cite{chainguard2023taming, darkreading2023false}, forcing analysts to manually inspect over 4,000 false alerts weekly~\cite{chainguard2023taming} and leading to alert fatigue that drives organizations to \wb{turn off} security tools or configure them permissively, creating exploitable security blind spots.

\begin{figure*}[htbp]
\centering
\includegraphics[width=\textwidth]{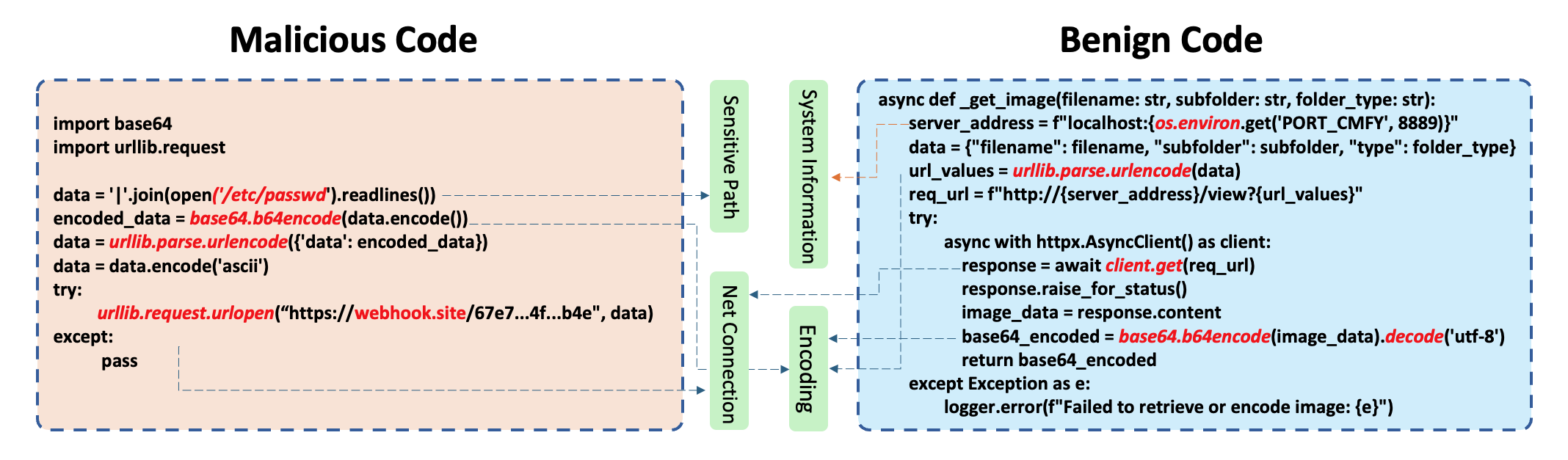}
\caption{API-level similarity between malicious and benign packages leading to false positive.}
\label{fig:motivation_example}
\end{figure*}

The fundamental limitation of existing detection tools lies in their reliance on coarse-grained, semantically unclear rules that fail to distinguish between benign and malicious behavioral boundaries, where rule-based static analysis methods employ simple pattern matching that captures broad suspicious behaviors without contextual nuances~\cite{chainguard2023hunting, guarddog2022}. For example, detecting network requests or file operations may flag legitimate packages performing similar actions for benign purposes~\cite{chainguard2023taming}, while existing approaches lack fine-grained knowledge about how identical operation sequences can serve entirely different purposes depending on execution context and intent~\cite{zhang2025killing}. This knowledge gap is particularly problematic as modern malware increasingly employs legitimate-looking API sequences and obfuscation techniques that mimic benign behaviors~\cite{eset2023malicious}, making detection systems unable to achieve the semantic-level precision required for practical deployment without a comprehensive understanding of contextual distinctions between legitimate and illegitimate behavioral patterns.

Our key insight is that detection failures themselves contain valuable knowledge that can be systematically extracted and leveraged to enhance malicious package identification, transforming the traditional reactive approach of simply reducing false positives into a proactive knowledge-driven detection enhancement framework. We propose a three-stage methodology: first, a comprehensive study phase that analyzes existing tools' false positives and false negatives on a curated dataset to extract fine-grained execution sequences and contextual patterns; second, a knowledge extraction phase that employs Large Language Models (LLMs) to dynamically generate semantic descriptions of program operations and systematically mine sequential patterns using gradient support thresholds to distinguish malicious from benign behaviors; and third, a RAG-enhanced detection phase that integrates extracted knowledge into a retrieval-augmented generation framework to provide context-aware reasoning for improved detection accuracy. The unique advantage of LLMs in this approach lies in their ability to understand semantic relationships between code contexts and their behaviors, enabling automated generation of fine-grained taxonomies and contextual knowledge that traditional rule-based systems cannot capture, thereby bridging the gap between low-level operation patterns and high-level semantic understanding required for precise malicious package detection.

This work makes four key contributions to malicious PyPI package detection: \textbf{(1)} We systematically analyze detection failures across multiple state-of-the-art tools on 18,137 PyPI packages, extracting 304 discriminative behavioral patterns that distinguish malicious from benign code through hierarchical pattern mining; \textbf{(2)} We develop an LLM-driven methodology that transforms concrete program operation sequences into semantic behavioral abstractions, enabling cross-implementation pattern recognition that transcends syntactic limitations; \textbf{(3)} We design a RAG-enhanced detection framework that achieves 99.50\% accuracy with only 2 false positives compared to 1,927-2,117 in existing tools, while maintaining 98.28\% accuracy on obfuscated code; \textbf{(4)} We demonstrate practical impact through real-world deployment, identifying 219 previously undetected malicious packages confirmed by PyPI officials and achieving effective cross-ecosystem generalization with 98.07\% accuracy on NPM packages.

\section{Preliminaries}
\label{sec:preliminaries}

We first introduce the problem definition and a motivating example of this work to demonstrate the key challenges in current malicious package detection.

\noindent \textbf{Problem Definition.} Current malicious package detection relies on static analysis to identify known threat signatures and behavioral patterns, particularly API calls, using either rule-based or learning-based pattern matching techniques. Formally expressed as:

\begin{equation}
is\_malicious(p) = \bigvee_{i=1}^{k} match(r_i, p)
\end{equation}
where $r \in R$ and $R$ is \wb{the set of matching rules}, $p$ represents a given suspicious package, if any rule $r_i$ matches the corresponding content, i.e., API calls, resource manipulations, or even suspicious names, then $p$ is considered malicious.

However, such designs make current malicious package detection techniques struggle with high false positive rates, mainly because of two key limitations: (1) they fail to achieve fine-grained analysis, often relying on coarse patterns like package names, metadata, or high-level code features that overlook subtle malicious behaviors; and (2) the patterns they use frequently lack discriminative power, making it difficult to distinguish malicious packages from benign ones that share similar structures or dependencies.

\noindent \textbf{Motivating Example.}
Consider two representative cases: the benign package \texttt{gen\_wrappers} (version 0.7.1), which performs legitimate image processing operations, and the malicious package \texttt{gmgeoip} (version 0.0.2), which exfiltrates sensitive system information to remote servers. Figure~\ref{fig:motivation_example} demonstrates that at the API level, both packages exhibit strikingly similar patterns: they both utilize \texttt{urllib.parse.urlencode} for URL parameter encoding and \texttt{base64.b64encode/decode} for data transformation. However, all leading state-of-the-art detection tools (i.e., Bandit4Mal, OSSGadget, and GuardDog) incorrectly flagged the benign package as malicious. After inspecting the detection logic of these tools, we found that they all classified \texttt{gen\_wrappers} (version 0.7.1) as malicious mainly because they simply \wb{classify} "shady-links" as malicious. 

This API-level similarity coupled with contradictory intentions demonstrates that, although current detection rules capture features of malicious code, they still fail to be determinative in distinguishing malicious intentions from benign ones in different contexts. To this end, in this paper, we aim to propose a detection framework that reduces false positives by incorporating a novel strategy to mine determinative pattern rules for \wb{malicious} code detection.

\section{Behavior Pattern Mining}
\label{sec:patternmining}

\begin{figure*}[t]
    \centering
    \includegraphics[width=1\textwidth]{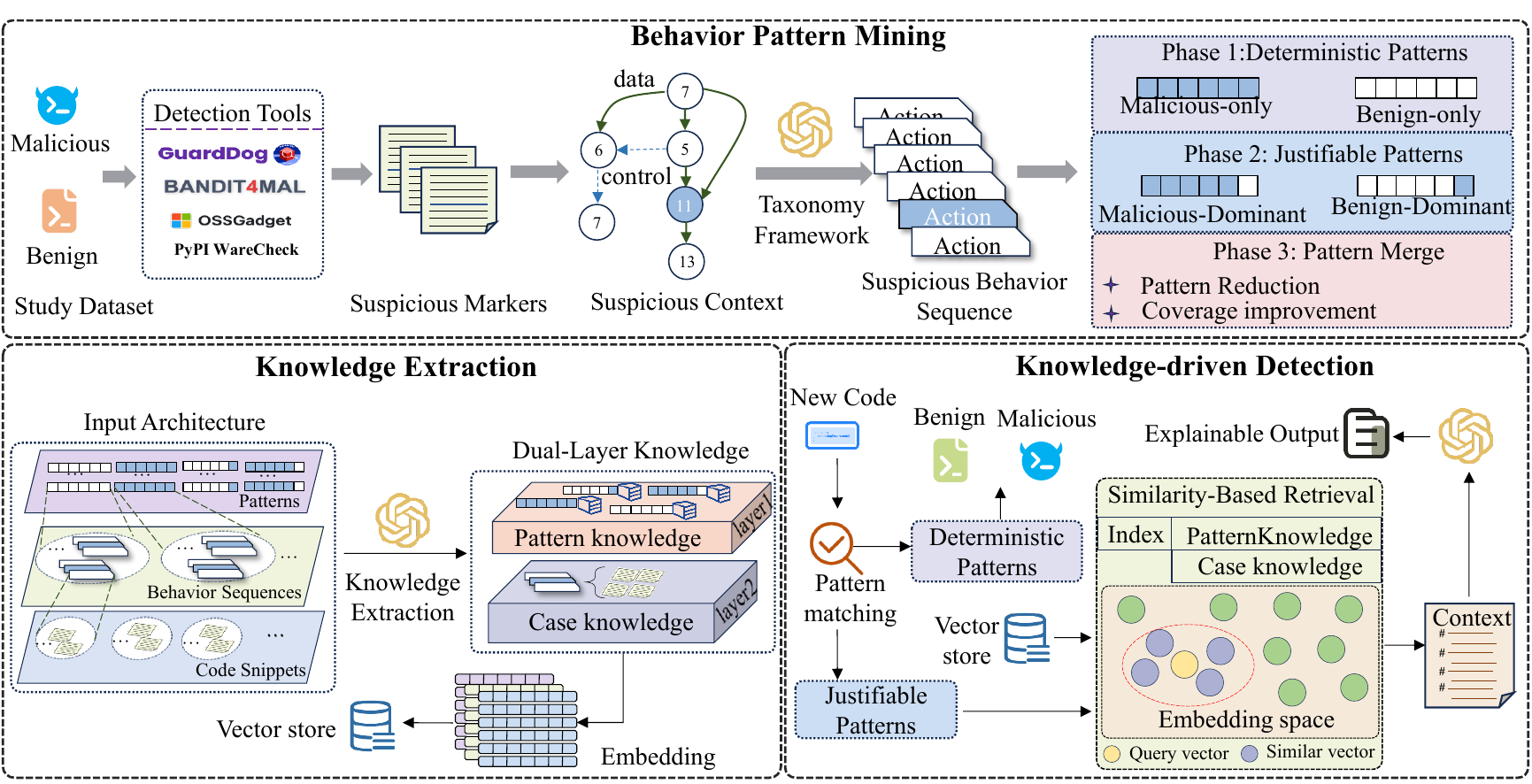}
    \caption{\wb{Overview} of Hierarchical Behavior Pattern Mining and Detection Framework}
    
    \label{fig:framework}
\end{figure*}

To this end, we propose a four-stage behavior pattern mining approach that systematically extracts behavioral patterns \wb{to distinguish between benign} and malicious packages at the semantic level. Specifically, we \ding{172} first collect suspicious codes that are either real malicious code or benign ones \wb{that can easily be classified} as malicious by existing tools, then, based on \ding{173} a well-constructed behavioral taxonomy, \ding{174} abstract their behaviors by corresponding actions and intentions behind, and finally, \ding{175} extract the representative subsequences that are discriminative and highly related to the classification of benign and malicious codes as behavioral patterns for further detection. Figure~\ref{fig:framework} illustrates the overview of this approach. 

\wb{We define the following key terms used throughout this paper:
\begin{itemize}
    \item \textbf{Action}: A semantic atomic behavioral operation representing a single meaningful step in code execution (e.g., \texttt{create\_socket}, \texttt{send\_http\_post}).
    \item \textbf{Action Sequence}: An ordered list of actions representing the execution flow of a code snippet.
    \item \textbf{Pattern}: Common subsequences shared across multiple action sequences that reveal fundamental differences between malicious and benign behaviors.
\end{itemize}}

\subsection{Suspicious Code Collection}
\label{sec:context_extraction}

Considering that identifying determinative patterns should be highly sensitive to the difference between benign code and malicious code, we first construct a dataset of suspicious code that \wb{is easily misclassified} by existing malicious code detection tools.

To this end, we employ four representative static analysis tools, GuardDog, Bandit4Mal, OSSGadget, and PyPI Malware Checks, as filters to identify potential malicious code. These tools are selected because they are open-source and static analysis-based tools, and they can produce detailed results containing the exact position of malicious code in user projects.
Based on them, for the well-selected 18K packages (half benign and half malicious, as detailed in~\Cref{subsec:study_dataset}), we 1) employ these tools to scan each of them to identify \wb{code snippets} that are considered malicious by these tools, and then 2) extract the related code context for each identified code snippet for further analysis.

\noindent \textbf{Suspicious Code Identification.} We employ these four tools to identify potential malicious code across the 18K packages. Each tool processes both benign and malicious packages, generating structured detection reports that label potential malicious code with their locations, line numbers, matched pattern types, and confidence scores (if any). Specifically, we parse each tool's scanning report for every package using regular expressions. We extract the complete file path information, line numbers, and code fragments identified by the tools \wb{to locate each suspicious code segment within the source files precisely}. For duplicate detections where multiple tools flag the same file location, we record only one instance since the surrounding code context remains identical.
Based on this, we can collect all true positives and false positives using these tools. After this, we also merged these results with all malicious code snippets from the ground truth (i.e., \wb{well-labeled} malicious code snippets for true malicious packages) to avoid missing false negatives, so that we can obtain a comprehensive dataset of suspicious code \wb{snippets that includes} all true malicious cases and false positive cases.

\noindent \textbf{Code Context Extraction.} Based on the collected suspicious codes, we extract the complete code context for each of them to understand their behavioral environment. For each identified code snippet, we read the entire Python source file and use the detection information (line numbers and code fragments) as clues to guide LLMs to extract their execution context, i.e., their data dependencies (variable definitions, assignments, and data flow leading to the flagged APIs) and control dependencies (conditional statements, loops, and function call chains that determine when and how the flagged code executes). This process ensures the extracted code segments are syntactically complete and preserve the logical structure necessary for understanding the operational environment surrounding each potential malicious code location. \wb{Note that the LLM in this phase serves purely as a context extraction tool, while the classification labels (malicious or benign) are determined by the ground truth dataset rather than LLM judgment.} The extraction prompt is available on our website~\cite{website}.

\begin{table}[t]
\centering
\caption{Taxonomy Categories and Subset of Behaviors}
\label{tab:api_classification}
\scalebox{0.66}{
\begin{footnotesize}
\begin{tabular}{ll||ll}
\toprule
\textbf{Category} & \textbf{Representative Behaviors} & \textbf{Category} & \textbf{Representative Behaviors} \\
\midrule
File Operations & basic\_file\_reading & Basic Network Ops & create\_http\_connection \\
& create\_directory & & send\_http\_post \\
& copy\_file & & establish\_tcp\_connection \\
\midrule
Network File Transfer & download\_file\_url & Command \& Control & execute\_shell\_command \\
& upload\_file\_transfer\_sh & & spawn\_process\_shell \\
& exfiltrate\_folder & & decrypt\_fernet\_data \\
\midrule
Third-party Platform & create\_discord\_bot & Data Exfiltration & init\_evil\_class \\
Abuse & send\_discord\_message & & init\_grabber\_class \\
& check\_roblox\_cookie & & init\_sender\_class \\
\midrule
Code Execution & exec\_python\_code & Info Gathering & get\_chrome\_passwords \\
& import\_dynamic & & capture\_screen\_region \\
& compile\_code\_object & & get\_os\_info \\
\midrule
Encryption/Hashing & generate\_aes\_cipher & System Operations & create\_child\_process \\
& decrypt\_aes\_data & & create\_thread \\
& create\_md5\_hash & & set\_registry\_value \\
\midrule
Data Transformation & decode\_base64\_to\_bytes & Persistence/Stealth & check\_persistence\_entry \\
& serialize\_to\_json & & open\_sqlite\_db \\
& convert\_int\_to\_char & & disable\_ssl\_warnings \\
\bottomrule
\end{tabular}
\end{footnotesize}
}
\footnotesize
\begin{enumerate}
    \item This table presents a subset of behavioral categories.
\end{enumerate}
\end{table}

\subsection{Context-aware Behavioral Modeling}\label{sec:api_taxonomy}

Next, we inspect the behavioral semantics of \wb{each identified code snippet to explore why it is considered} malicious using existing tools, based on the extracted execution context. 

Before summarizing the behavioral semantics, we first develop a behavioral taxonomy to describe the execution behaviors of code (i.e., intentions of code snippets) for better understanding and presentation. 
Specifically, for the identified context of each suspicious code snippet, we leverage the LLMs to incorporate the Card Sorting strategy to generate summary text (i.e., cards) to summarize the behaviors and potential intentions in their context for each identified action (i.e., operations, such as declarations, assignments, and function calls). In detail, there are three major steps:

\noindent \textbf{Behavioral Summarization.} We first summarize execution behavior for each identified action. Considering that obfuscation techniques, such as alias imports, dynamic imports, and nested function calls, are often employed in malicious packages to evade detection, we first incorporate LLMs to identify and resolve these obfuscation patterns while preserving actual execution order to de-obfuscate code snippets. Next, we ask LLMs to generate descriptions of corresponding behaviors within 20 words for each identified action, during which LLMs are required to neutrally summarize the behavior of the specific action in the context of a given code snippet. Based on this, we can tag each identified action with descriptive text under its corresponding context.

\noindent \textbf{Card Sorting.} Considering that LLMs could use different descriptive text to summarize similar actions and intentions, we employ the hybrid Card Sorting strategy during the summarization. Specifically, for the set of actions we identified, we randomize them into \wb{three different ordered lists}, and we ask the LLMs to process them in sequence. For each action during the summarization of a sequence, apart from requiring the LLMs to do the summarization (generate a card), we also provide the LLMs with a collected set of cards (generated summary texts of previous actions), and ask the LLMs to choose these existing ones if suitable, before generating a new one. Based on this, we are able to obtain three generated card sets after the automated summarization.

\noindent \textbf{Behavioral Taxonomy Construction.} Since the LLMs could introduce uncertainty and hallucination during the summarization, we also conduct human-guided consolidation using three domain experts (two doctoral candidates and one postdoc, with at least three years of experience in malicious code detection) to independently review all generated cards, exclude nonsensical text, and merge semantically similar descriptions into unified categories. When at least two experts approve a decision, \wb{we proceed with the corresponding action} to refine the generated cards. Based on this, we can generate a set of summary texts for the behaviors, actions, and their intentions in identified suspicious code snippets.

Our results showed that, \wb{from the 8,543 malicious packages, we extracted 2,283 unique malicious code snippets after deduplication. This reduction occurs because malicious packages are often released in batches with identical malicious code embedded. From these snippets,} we identified 13,834 actions in total, which were reduced to 837 distinct ones after deduplication. The LLM-driven semantic analysis generated 495 initial descriptions, which were consolidated through expert review into a refined taxonomy of 327 unique API behavioral categories. A subset of categories is shown in Table~\ref{tab:api_classification}.

\subsection{Context-to-Action Sequence Mapping}
\label{sec:action_sequence}

With the code context extracted from Section~\ref{sec:context_extraction} and the behavioral taxonomy constructed in Section~\ref{sec:api_taxonomy}, we then abstract the specific code implementations into behavioral sequences that capture the essential execution logic. This abstraction process converts specific operations in code into generalized behavioral sequences, where each item in the sequence represents a discrete step toward achieving a broader behavioral objective. For instance, code implementations may use different APIs like \texttt{urllib.request} or \texttt{requests.get}, \wb{both of which map} to the abstract action "network\_communication" \wb{within data leak patterns}.

\subsection{Hierarchical Pattern Discovery}

\wb{After the abstraction of behavioral sequences for each code snippet, we then identified the major behavioral patterns that can reliably distinguish malicious code behaviors from benign ones.}
To this end, we designed a systematic process to hierarchically discover two specific types of patterns, \ding{172} deterministic patterns that provide definitive classification signals, and \ding{173} justifiable patterns that require contextual reasoning for accurate assessment.

Our pattern discovery approach employs a hierarchical mining strategy using the PrefixSpan algorithm to systematically identify behavioral subsequences (i.e., identified patterns) that are recognizable with the classification of benign and malicious packages. The core methodology involves iterative identification of such subsequence patterns, where each iteration focuses on discovering subsequences that were not captured in previous rounds with progressively decreasing support thresholds, which ensures that both common and rare but significant behavioral patterns can be systematically identified. The detailed process is presented in Algorithm~\ref{algorithm_hierarchical}.

\noindent \textbf{Phase 1: Deterministic Pattern Mining.} We first identify patterns with perfect classification reliability by extracting patterns that appear exclusively in either benign or malicious action sequences. 
\wb{Let $R_B$ and $R_M$ denote the sets of action sequences extracted from benign and malicious code snippets, respectively.} For a given predefined support value $s \in S$, we apply the PrefixSpan algorithm on the combined dataset $R_B \cup R_M$ to discover frequent subsequences.
For each discovered pattern $p$, we check whether it covers only benign action sequences or only malicious action sequences. Patterns that exclusively cover benign sequences are collected into $det_B = \{p \mid p \in patterns \wedge \text{CoverOnly}(p, R_B)\}$, while patterns that exclusively cover malicious sequences form $det_M = \{p \mid p \in patterns \wedge \text{CoverOnly}(p, R_M)\}$. We combine these exclusive patterns into our deterministic pattern set using $P_{det} \leftarrow P_{det} \cup det_B \cup det_M$. 
After each round of mining, we remove all sequences covered by the newly discovered deterministic patterns from our working datasets to ensure only uncovered action sequences are mined in subsequent rounds: $R_B \leftarrow R_B \setminus \text{CoveredBy}(det_B \cup det_M)$ and $R_M \leftarrow R_M \setminus \text{CoveredBy}(det_B \cup det_M)$. This process repeats across all predefined support values, ensuring that each deterministic pattern provides 100\% classification confidence.

\noindent \textbf{Phase 2: Justifiable Pattern Mining.} Next, we target the remaining uncovered sequences in $R_B$ and $R_M$ that could not be perfectly classified by these identified deterministic patterns, where we aim to further identify patterns with strong but imperfect discriminative power. In this phase, we still apply the PrefixSpan algorithm on the residual datasets $R_B \cup R_M$ at each support level $s \in S$ to discover patterns from the remaining behavioral sequences that escaped deterministic classification. For each discovered pattern $p$, we calculate its classification bias using $MaxCoverageRatio(p, R_B, R_M)$, which computes the percentage of action sequences covered by pattern $p$ that belong to the dominant class. For example, if pattern $p$ covers 10 action sequences where 9 are malicious and 1 is benign, the ratio is 0.9, indicating 90\% bias toward malicious behavior. We select patterns with strong bias by checking if the ratio meets our threshold: when $ratio \geq \tau$ (set to \wb{0.9 empirically}), we add the pattern to our justification set using \wb{$P_{just} \leftarrow P_{just} \cup \{p\}$}. This process identifies patterns that, while not providing perfect separation, still offer valuable probabilistic classification signals for action sequences that resist deterministic classification. This also follows the iterative mining process by the predefined order of support value $s \in S$.

\noindent \textbf{Phase 3: Pattern Merge.} After the identification, there is a pattern explosion problem that PrefixSpan \wb{could generate} numerous overlapping sub-patterns from each action sequence, making the pattern set impractical for deployment due to excessive redundancy. Therefore, we employ a greedy algorithm to select the minimal pattern subset from \wb{$P_{det} \cup P_{just}$} that achieves complete action sequence coverage. Starting with an empty result set $P_{opt}$ and an empty $covered$ set to track which action sequences have been covered, the algorithm performs iterative selection. In each iteration, we evaluate all remaining patterns in \wb{$P_{det} \cup P_{just}$} to find the one that covers the maximum number of previously uncovered action sequences, calculated as $|\text{CoveredBy}(p) \setminus covered|$ for each pattern $p$. We select the pattern with the highest coverage using $best\_p \leftarrow \arg\max_{p \in (P_{det} \cup P_{just})} |\text{CoveredBy}(p) \setminus covered|$, add this selected pattern to our final set $P_{opt}$, and update our tracking set by adding all action sequences covered by this pattern: $covered \leftarrow covered \cup \text{CoveredBy}(best\_p)$. This selection process repeats until either all action sequences are covered or no remaining pattern can increase coverage, ensuring we obtain the minimum number of patterns required for complete behavioral coverage.

After these three steps of actively mining deterministic and justifiable patterns, \wb{we identified the corresponding patterns that can be easily used to classify malicious and benign code snippets. These patterns serve as the knowledge base for downstream malicious code detection.}

\begin{algorithm}[t]
\centering
\scriptsize
\scalebox{0.9}{%
\begin{minipage}{1.25\linewidth}
\label{algorithm_hierarchical}
\SetAlgoLined
\SetKwFunction{FExtractSequences}{ExtractSequences}
\SetKwFunction{FPrefixSpan}{PrefixSpan}
\SetKwFunction{FCoverOnly}{CoverOnly}
\SetKwFunction{FCoveredBy}{CoveredBy}
\SetKwFunction{FMaxCoverageRatio}{MaxCoverageRatio}
\SetKwFunction{FAllIndices}{AllIndices}
\SetKwInOut{Input}{input}
\SetKwInOut{Output}{output}
\Input{Benign samples $B$, Malware samples $M$, Support \\ levels $S = \{s_1,s_k\}$,  Distinction threshold $\tau$}
\Output{Merged pattern set  $P_{opt}$}
$B_{seq} \leftarrow$ \FExtractSequences{$B$}\;
$M_{seq} \leftarrow$ \FExtractSequences{$M$}\;
$R_B \leftarrow B_{seq}$; $R_M \leftarrow M_{seq}$; $P_{det} \leftarrow \emptyset$\;
\BlankLine
\tcp{Phase 1: Deterministic pattern mining}
\ForEach{$s \in S$}{
    $patterns \leftarrow$ \FPrefixSpan{$R_B \cup R_M, s$}\;
    $det_B \leftarrow \{p \mid p \in patterns \wedge$ \FCoverOnly{$p, R_B$}$\}$\;
    $det_M \leftarrow \{p \mid p \in patterns \wedge$ \FCoverOnly{$p, R_M$}$\}$\;
    $P_{det} \leftarrow P_{det} \cup det_B \cup det_M$\;
    $R_B \leftarrow R_B \setminus$ \FCoveredBy{$det_B \cup det_M$}\;
    $R_M \leftarrow R_M \setminus$ \FCoveredBy{$det_B \cup det_M$}\;
}
\BlankLine
\tcp{Phase 2: Justification pattern mining}
$P_{just} \leftarrow \emptyset$\;
\ForEach{$s \in S$}{
    $patterns \leftarrow$ \FPrefixSpan{$R_B \cup R_M, s$}\;
    \ForEach{$p \in patterns$}{
        $ratio \leftarrow$ \FMaxCoverageRatio{$p, R_B, R_M$}\;
        \If{$ratio \geq \tau$}{
            $P_{just} \leftarrow P_{just} \cup \{p\}$\;
        }
    }
}
\BlankLine
\tcp{Phase 3: Pattern Merge}
$P_{opt} \leftarrow \emptyset$; $covered \leftarrow \emptyset$\;
$all\_indices \leftarrow$ \FAllIndices{$B_{seq}, M_{seq}$}\;
\While{$covered \neq all\_indices$}{
    $best\_p \leftarrow \arg\max_{p \in (P_{det} \cup P_{just})} |$\FCoveredBy{$p$}$ \setminus covered|$\;
    \If{$|$\FCoveredBy{$best\_p$}$ \setminus covered| > 0$}{
        $P_{opt} \leftarrow P_{opt} \cup \{best\_p\}$\;
        $covered \leftarrow covered \cup$ \FCoveredBy{$best\_p$}\;
    }
    \Else{
        \textbf{break}\;
    }
}
\BlankLine
\Return{$P_{opt}$}\;
\caption{Hierarchical Sequence Pattern Mining}
\end{minipage}
}
\end{algorithm}
\section{Study}
\label{sec:study}

We first conducted an empirical investigation to understand the distribution, coverage, and potential applicability of these extracted behavioral patterns on malicious code detection by answering the following research question: 

\noindent \textbf{RQ1. Distribution Analysis:} What behavioral patterns emerge from hierarchical mining and how suitable are they for detection rules?

\subsection{Dataset Construction}
\label{subsec:study_dataset}

We constructed a large-scale dataset of PyPI packages to support systematic analysis of malicious behavior patterns. The dataset consists of 9,552 unique packages across 10,906 versions, sourced from the curated collection by Guo et al.\cite{guo2023empirical}, which provides confirmed ground-truth labels for supply chain attacks. 
\wb{For benign packages, we selected 11,988 packages based on PyPI 
download statistics from the preceding month. We prioritized highly 
downloaded packages because prior studies have shown that popular 
packages trigger significantly higher false positive rates than 
randomly sampled packages~\cite{vu2023bad, darkreading2023false}, 
providing a more rigorous test for detection precision. Additionally, 
popular packages are less likely to contain malicious code, as any 
malicious behavior would be quickly discovered given their large user 
base. These packages also represent real-world scenarios with higher 
code complexity and functional diversity. We verified with mainstream 
vulnerability databases (OSV~\cite{osv2024} and Snyk) that no known 
malicious packages were present in our benign dataset.  
For each package, we selected one random version to avoid redundancy, as 
different versions of the same package exhibit minimal 
code differences. We added a 5-second delay between downloads to 
minimize server load. Our one-time collection of approximately 
12,000 packages represents less than 0.0006\% of PyPI's 2.1 billion~\cite{pypi_stats_2024}
daily downloads. We did not collect any personally 
identifiable information during this process.
We partitioned the complete dataset using an 80-20 split, 
allocating 18,137 packages for pattern analysis and reserving 
4,757 for unbiased evaluation.}

\subsection{Experimental Setup}

\begin{table}[t]
\centering
\caption{Detection statistics for static analysis tools}
\label{tab:detection_statistics}
\scriptsize
\begin{tabular}{lcccc}
\toprule
\textbf{Tool} & \textbf{TP} & \textbf{FP} & \textbf{FN} & \textbf{Detection Positions} \\
\midrule
Bandit4Mal     & 3,899 & 7,699 & 4,633 & 695,974 \\
GuardDog       & 6,640 &   447 &   757 &  11,420 \\
OSSGadget      & 7,529 & 8,113 & 1,001 & 4,955,383 \\
PyPI Warehouse & 8,332 & 8,422 &   208 &   248,737 \\
\bottomrule
\end{tabular}
\end{table}

We applied four static analysis tools to all 18,137 packages in the study partition to identify potential locations of malicious code. \wb{These tools exhibit low false negative rates, with only 64 out of 8,543 malicious packages being missed by all four tools simultaneously. Our taxonomy captures behavioral intentions rather than specific code implementations. Different implementations often share the same malicious intent, so these missed packages likely do not introduce new behavioral patterns.} As shown in Table~\ref{tab:detection_statistics}, the tools exhibit varying detection characteristics: GuardDog identified 11,420 positions, while OSSGadget flagged 4,955,383 positions, demonstrating different sensitivity thresholds. After deduplication of overlapping detections at identical file locations, we obtained 956,131 unique detection positions for subsequent context extraction.

For each Python file containing detection positions, we consolidate all flagged locations within the file and employ GPT-4.1 to extract comprehensive code snippets in a single analysis pass. Through this process, we successfully collected 5,813 false positive contexts (benign code incorrectly flagged as malicious) and 7,458 malicious code snippets from actual malicious and benign packages.

From these 13,271 code snippets, we identified 13,834 individual API calls. \wb{Using card sorting with GPT-4.1, 495 initial behavioral descriptions were generated at a cost of 126 USD. Three domain experts then reviewed these descriptions and merged semantically similar ones into 327 unique behavioral categories, as shown in Table~\ref{tab:api_classification}. This taxonomy construction is a one-time effort.}

We mapped the extracted code snippets to abstract action sequences using the constructed API taxonomy, generating 2,431 action sequences with an average length of 4.2 actions per sequence. The action sequences comprise 1,350 benign sequences and 1,081 malicious sequences. For the three-phase \wb{hierarchical} mining process, we configure support thresholds as $S = \{30, 25, 20, 15, 10, 7, 5, 3, 2\}$ and distinction threshold $\tau = 0.9$. Phase 1 deterministic pattern mining discovered 115,960 pure patterns achieving 72.44\% overall coverage. Phase 2 justification pattern mining processes remaining uncovered sequences and identified 47 patterns with $\geq 90\%$ class bias, contributing additional 23.41\% coverage. The combined phases produced 116,007 patterns, covering 95.85\% of action sequences. Phase 3 pattern merge applied greedy set cover optimization, reducing the pattern set from 116,007 to 304 patterns (a 99.74\% reduction) while maintaining 92.60\% sequence coverage. The final optimized pattern set contains 278 deterministic patterns and 26 justifiable patterns.




\subsection{Pattern Extraction Evaluation}

\wb{To validate the quality of action sequence extraction, we randomly selected 500 code snippets covering both malicious and benign packages. Three independent security experts were provided with the original code snippets and the taxonomy of actions, and we asked the experts to manually label (1) the malicious behavior related actions implemented in the code snippets, and (2) the action sequence that constitutes the complete malicious behavior process. For both of them, we adopted majority voting after they discuss when inconsistencies occur.}

\wb{As for the results, at the action level, we retrieved 3,608 individual actions from the 500 code snippets. Among these, 3,432 actions (95.12\%) were correctly mapped to their semantic labels. The errors consisted primarily of hallucinated actions (174, 4.82\%) where the model generated actions not present in the code, and wrong-type mappings (2, 0.06\%) where similar operations were mislabeled. Notably, no missing actions were reported, indicating the extraction tends toward over-generation rather than under-generation.}

\wb{At the sequence level, 461 out of 500 snippets (92.2\%) were correctly extracted with all actions accurately identified. The remaining 39 snippets contained errors, primarily including hallucinated actions, over-abstraction, and incorrect semantic mapping. The evaluation achieved expert consensus on 97.2\% of samples with a Fleiss' Kappa~\cite{fleiss1971measuring} coefficient of 0.8778, indicating almost perfect inter-rater agreement, which validates the reliability of our evaluation protocol.}

\subsection{RQ1. Distribution Analysis}

To understand the practical value of our extracted patterns, we first investigated the distribution, coverage, and complexity of these identified patterns to evaluate their applicability in real-world malicious code detection.

\noindent \textbf{Pattern Distribution.} The final pattern set contains 278 deterministic patterns (91.4\%) and 26 justifiable patterns (8.6\%), indicating that the majority of extracted behavioral signatures provide definitive classification signals. Among the deterministic patterns, 192 patterns (69.1\%) exclusively identify benign behaviors, while 86 patterns (30.9\%) exclusively cover malicious behaviors. This 2.2:1 ratio reflects the diversity of legitimate functionalities compared to concentrated malicious attack vectors. The 26 justifiable patterns exhibit varying degrees of class bias, with 12 patterns showing strong benign dominance ($>80\%$ benign coverage), 11 patterns demonstrating mixed characteristics (50-80\% bias), and 3 patterns exhibiting malicious dominance ($>80\%$ malicious coverage).

\noindent \textbf{Coverage and Discriminative Power.} Deterministic patterns demonstrate superior coverage capabilities, with malicious-only patterns achieving 87.7\% coverage of malicious action sequences (948/1,081) while benign-only patterns cover 60.2\% of benign sequences (813/1,350). This asymmetric coverage indicates that malicious behaviors exhibit more concentrated and predictable patterns compared to the diverse legitimate functionalities in benign packages. Justifiable patterns provide crucial coverage for boundary cases, covering an additional 200 action sequences that could not be definitively classified, representing 8.2\% of the total sequence space.

\noindent \textbf{Pattern Complexity and Behavioral Characteristics.} Analysis of pattern lengths reveals that deterministic patterns tend to be more specific, with malicious-only patterns averaging 2.8 actions per pattern, while benign-only patterns average 3.2 actions. The most frequent malicious patterns include complex attack sequences such as [\texttt{create\_socket, establish\_tcp\_connection, dup\_socket\_stdin, dup\_socket\_stdout, dup\_socket\_stderr}] covering 29 malicious sequences for reverse shell establishment, and [\texttt{get\_env\_var, get\_clipboard\_text, copy\_to\_clipboard}] appearing in 43 sequences for information harvesting. In contrast, benign patterns demonstrate functional diversity with legitimate sequences like [\texttt{get\_env\_var, spawn\_process\_no\_shell, read\_process\_stdout}] for system administration tasks and [\texttt{path\_string\_operations, execute\_shell\_command, exit\_program}] for package installation processes. Justifiable patterns exhibit intermediate complexity, typically involving 2-4 actions that require contextual analysis to determine intent, such as patterns combining file operations with network communications that could serve either legitimate or malicious purposes.

\responsebox{Response to RQ1: Our hierarchical mining approach successfully extracts 304 discriminative behavioral patterns, with 278 deterministic patterns providing definitive classification signals and 26 justifiable patterns handling boundary cases. The patterns demonstrate clear behavioral distinctions, achieving 92.60\% overall sequence coverage.}
\section{\tool{}}
\label{sec:methodology}

We transform the behavioral patterns from our empirical analysis into an operational detection system. 
Based on the identified behavioral patterns that can be informative to classify benign and malicious code, we further implement a classification system (\tool{}) to demonstrate their efficacy.
The system bridges static pattern discoveries (i.e., the 278 deterministic patterns enabling direct classification and the 26 justification patterns that provide strong indications) with contextual reasoning, by incorporating RAG-enhanced analysis for greater generalizability. In this section, we introduce the design of \tool{}, including \ding{172} the construction of knowledge base for RAG, and \ding{173} the working pipeline of \tool{} to detect malicious packages. 
Figure~\ref{fig:framework} illustrates the overview of \tool{}.

\subsection{Knowledge Base Construction}
\label{sec:ragdb}
The knowledge base transforms behavioral patterns from the mining phase into structured detection knowledge through systematic extraction and indexing. Let $\mathcal{P} = P_{det} \cup P_{just}$ denote patterns from Section~\ref{sec:patternmining}, where $P_{det}$ contains 278 deterministic patterns and $P_{just}$ contains 26 justification patterns.

\noindent \textbf{Dual-Layer Knowledge Storage.} The pattern knowledge layer stores each pattern $p \in \mathcal{P}$ with its semantic interpretation and classification properties. For deterministic patterns where $p \in P_{det}$, we annotate consistent behavioral characteristics: attack vectors for malicious-only patterns and legitimate use cases for benign-only patterns. For justification patterns where $p \in P_{just}$, we extract distinction rules that capture contextual differences determining classification. The case knowledge layer preserves the mapping between patterns and their original implementations. Each pattern $p$ links to its covered action sequences $S_p = \{s \mid s \text{ is covered by } p\}$ and corresponding code snippets. We pre-compute embedding vectors for all sequences and contexts using \wb{text-embedding-3-large~\cite{openai2024embedding}}, producing $e(s) \in \mathbb{R}^{3072}$ for each sequence, enabling efficient similarity search during detection.

\noindent \textbf{Knowledge Extraction Process.} For each pattern $p \in \mathcal{P}$, we employ LLM to analyze its constituent sequences and extract detection-relevant insights. For deterministic patterns where $p \in P_{det}$, the LLM synthesizes common characteristics across all implementations in $S_p$. For justification patterns where $p \in P_{just}$, we partition the covered sequences into $S_p^B$ (benign instances) and $S_p^M$ (malicious instances). The LLM conducts comparative analysis between $S_p^B$ and $S_p^M$, identifying contextual factors that differentiate them, such as data flow destinations (user files versus system credentials), network endpoints (local versus external), and execution triggers (user-initiated versus automated). These insights produce distinction rules that guide classification when the same behavioral pattern appears in different contexts.

\noindent \textbf{Knowledge Indexing.} The extracted knowledge undergoes multi-modal indexing to support different retrieval strategies. We implement hash-based indexing for exact pattern matching, subsequence indexing for partial pattern detection, and FAISS vector indexing for similarity search across all case embeddings. This ensures efficient retrieval whether input sequences match patterns exactly, partially, or semantically.

\subsection{Detection Pipeline}
\label{sec:ragdetection}

The detection pipeline applies the constructed knowledge base to classify unknown packages through behavioral analysis and contextual reasoning. Given an unknown package, we extract behavioral sequences using the same methodology from Section~\ref{sec:patternmining}, then leverage the knowledge base to determine classification.

\noindent \textbf{1) Behavioral Extraction and Pattern Matching.} For each Python file in an unknown package, we identify sensitive APIs from our category taxonomy and extract their execution contexts using the same LLM-based approach from training. The APIs are mapped to behavioral categories, generating action sequence $s_{new}$. We first attempt exact and subsequence matching against patterns in $\mathcal{P}$. If $s_{new}$ matches a deterministic pattern $p \in P_{det}$, classification is immediate: malicious-only patterns flag the package as malicious while benign-only patterns indicate legitimate functionality. \wb{If $s_{new}$ matches both benign and malicious deterministic patterns, the file is classified as malicious since any malicious pattern indicates a potential threat.}

\noindent \textbf{2) Similarity-Based Retrieval.} When \wb{$s_{new}$ does not match any deterministic pattern, or when it} yields only justification patterns $p \in P_{just}$, we employ similarity-based retrieval to find relevant cases. 
We compute embedding $e(s_{new})$ for the input sequence and separately calculate behavioral similarity $\text{sim}_s(e(s_{new}), e(s_i))$ and code context similarity $\text{sim}_c(c_{new}, c_i)$ with all stored sequences in $S_p$. We retrieve the top-k \wb{(k=5)} most similar cases based on behavioral similarity and the top-k most similar cases based on context similarity from both $S_p^B$ and $S_p^M$, combining these retrieved sets to provide comprehensive similarity coverage for subsequent RAG-enhanced analysis.

\noindent \textbf{3) Knowledge-driven Detection.} For justification patterns, classification requires contextual reasoning. The LLM receives: (1) the target code $c_{new}$ and its sequence $s_{new}$, (2) the matched pattern $p$ with its distinction rules, (3) top-k similar benign cases from $S_p^B$ with similarity scores, and (4) top-k similar malicious cases from $S_p^M$ with similarity scores. The LLM analyzes whether $c_{new}$ satisfies the distinction rules, comparing against retrieved examples to determine behavioral intent. For instance, if the distinction rule indicates that "base64 encoding of system files suggests malicious intent while encoding user documents is benign," the model examines the data sources in $c_{new}$. The output provides classification $f(s_{new}) \in \{\text{benign}, \text{malicious}\}$, confidence score based on evidence strength, and explanatory reasoning tracing specific rules and examples that influenced the decision.
\section{Evaluation}
\label{sec:evaluation}

To comprehensively validate our approach, we conducted systematic experiments addressing four interconnected research questions. Our evaluation covers detection accuracy comparison with state-of-the-art tools, ablation analysis of knowledge components, real-world deployment effectiveness, and cross-ecosystem generalizability.

\noindent \textbf{RQ2. Accuracy}: How does our knowledge-driven framework compare to SOTA tools?

\noindent \textbf{RQ3. Ablation}: What is the impact of knowledge integration on framework performance?

\noindent \textbf{RQ4. Usability}: How effective is the \tool{} against emerging threats in real-world scenarios?

\noindent \textbf{RQ5. Cross-ecosystem}: How effective is the extracted knowledge when applied to other package ecosystems?

\begin{table}[htbp]
\centering
\caption{Evaluation Dataset Characteristics}
\label{tab:eval_datasets}
\scriptsize
\begin{tabular}{lcccl}
\toprule
\textbf{Dataset} & \textbf{Benign} & \textbf{Malicious} & \textbf{Total} & \textbf{Ecosystem}\\
\midrule
Original Test Packages   & 2,394 & 2,363 & 4,757 & PyPI \\
Latest Packages  & 1,001 & 1,097 & 2,098 & PyPI \\
Obfuscated Packages    & 1,019 & 1,050 & 2,069  & PyPI \\
NPM Packages    & 953 & 809 & 1,762  & NPM \\
\midrule
\textbf{Total}  & \textbf{4,414} & \textbf{4,510} & \textbf{8,924}  & / \\
\bottomrule
\end{tabular}
\end{table}

\begin{table*}[h]
\centering
\caption{Comparison of Baseline Tools for Malicious Package Detection}
\label{tab:baseline_selection}
\scalebox{0.8}{
\begin{footnotesize}
\begin{tabularx}{1.15\textwidth}{l>{\raggedright\arraybackslash}X|cccc|cc|l>{\raggedright\arraybackslash}X|cccc|cc}
\toprule
\multirow{2}{*}{\textbf{Baseline}} & \multirow{2}{*}{\textbf{Tech.}} & \multicolumn{4}{c|}{\textbf{Output}} & \multirow{2}{*}{\textbf{Avail.}} & \multirow{2}{*}{\textbf{Date}} & 
\multirow{2}{*}{\textbf{Baseline}} & \multirow{2}{*}{\textbf{Tech.}} & \multicolumn{4}{c|}{\textbf{Output}} & \multirow{2}{*}{\textbf{Avail.}} & \multirow{2}{*}{\textbf{Date}} \\
\cmidrule(lr){3-6}\cmidrule(lr){11-14}
& & \textbf{MB} & \textbf{BP} & \textbf{CL} & \textbf{KC} & & & & & \textbf{MB} & \textbf{BP} & \textbf{CL} & \textbf{KC} & & \\
\midrule
\addlinespace[0.5ex]
Bandit4Mal~\cite{bandit4mal2024}& Rule & \ding{51} & \ding{51} & \ding{51} & \ding{53} & \ding{51} & 2022 & 
OSSGadget~\cite{ossgadget2024} & Rule & \ding{51} & \ding{51} & \ding{51} & \ding{53} & \ding{51} & 2025 \\[0.5ex]
SocketAI~\cite{zahan2025leveraging} & LLM & \ding{51} & \ding{51} & \ding{51} & \ding{53} & \ding{51} & 2025 & 
PyPI WareCheck~\cite{warehouse2020malware} & Rule & \ding{51} & \ding{51} & \ding{51} & \ding{53} & \ding{51} & 2024 \\[0.5ex]
MalOSS~\cite{duan2020towards} & Dynamic & \ding{51} & \ding{53} & \ding{51} & \ding{53} & \ding{53} & 2021 & 
Guarddog~\cite{Guarddog2023} & Rule & \ding{51} & \ding{51} & \ding{51} & \ding{53} & \ding{51} & 2025 \\[0.5ex]
Cerebro~\cite{zhang2025killing} & ML & \ding{51} & \ding{51} & \ding{53} & \ding{53} & \ding{51} & 2025 & 
SAP~\cite{ladisa2023feasibility} & ML & \ding{53} & \ding{53} & \ding{53} & \ding{53} & \ding{51} & 2024 \\[0.5ex]
ClamAV~\cite{clamav2024} & ML & \ding{53} & \ding{53} & \ding{53} & \ding{53} & \ding{53} & 2024 & 
MalWuKong~\cite{li2023malwukong} & Rule & \ding{51} & \ding{51} & \ding{51} & \ding{53} & \ding{53} & 2023 \\[0.5ex]
OSCAR~\cite{zheng2024towards} & Dynamic & \ding{51} & \ding{51} & \ding{51} & \ding{53} & \ding{53} & 2024 & 
MPHunter~\cite{liang2023needle} & ML & \ding{53} & \ding{53} & \ding{53} & \ding{53} & \ding{53} & 2023 \\
Hercule~\cite{shariffdeen2025detecting} & Dynamic & \ding{51} & \ding{51} & \ding{51} & \ding{53} & \ding{51} & 2025 & 
MalGuard~\cite{gao2025malguard} & ML & \ding{51} & \ding{53} & \ding{53} & \ding{53} & \ding{51} & 2025 \\
\addlinespace[0.5ex]
\bottomrule
\end{tabularx}
\end{footnotesize}
}
\\[4pt]  
\footnotesize
\textit{\textbf{Tech.}} refers to the technology used by the baseline. In \textbf{Output}: \textit{\textbf{MB}} (Malicious Behavior), \textit{\textbf{BP}} (Behavior Pattern), \textit{\textbf{CL}} (Code Location), \textit{\textbf{KC}} (Key Context). \textit{\textbf{Avail.}} indicates the availability of the baseline, where \ding{51} denotes availability and \ding{53} indicates unavailability. \textit{\textbf{Date}} is the most recent update time.
\end{table*}

{
\setlength{\tabcolsep}{1.85pt}

\begin{table}[!t]
\centering
\caption{Performance of \tool{} and Baseline Tools}
\label{tab:rq2_results}
\scriptsize
\begin{tabular}{llrrrrrr}
\toprule
\textbf{Dataset} & \textbf{Tool} & \textbf{Accuracy} & \textbf{Precision} & \textbf{Recall} & \textbf{F1-Score} & \textbf{FP} & \textbf{FN} \\
\midrule
\multirow{8}{*}{\makecell[l]{Evaluation \\Dataset}}
& Bandit4Mal & 33.37\% & 34.72\% & 48.69\% & 40.54\% & 1,927 & 1,080 \\
& GuardDog & 94.12\% & 92.38\% & 94.42\% & 93.39\% & 141 & 101 \\
& OSSGadget & 50.01\% & 48.13\% & 88.38\% & 62.32\% & 2,000 & 244 \\
& PyPI WareCheck & 52.99\% & 49.66\% & 99.19\% & 66.18\% & 2,117 & 17 \\
& SAP-DT & 59.99\% & 68.68\% & 25.42\% & 37.10\% & 244 & 1,570 \\
& SAP-RF & 86.74\% & 92.20\% & 78.05\% & 84.54\% & 139 & 462 \\
& SAP-XGB & 65.55\% & 80.00\% & 34.39\% & 48.11\% & 181 & 1,381 \\
& \wb{Cerebro} & \wb{89.12\%} & \wb{90.21\%} & \wb{85.88\%} & \wb{88.00\%} & \wb{196} & \wb{297} \\
& \wb{Hercule} & \wb{88.76\%} & \wb{90.67\%} & \wb{84.47\%} & \wb{87.46\%} & \wb{183} & \wb{327} \\
& SocketAI & 90.95\% & 92.01\% & 88.12\% & 90.03\% & 161 & 250 \\
& \textbf{PyGuard} & \textbf{99.50\%} & \textbf{99.90\%} & \textbf{99.08\%} & \textbf{99.49\%} & \textbf{2} & \textbf{19} \\
\midrule
\multirow{8}{*}{\makecell[l]{Latest \\ Dataset}}
& Bandit4Mal & 58.45\% & 49.60\% & 94.40\% & 65.03\% & 822 & 48 \\
& GuardDog & 83.84\% & 89.06\% & 68.12\% & 77.20\% & 69 & 263 \\
& OSSGadget & 52.63\% & 45.79\% & 86.33\% & 59.84\% & 875 & 117 \\
& PyPI WareCheck & 59.87\% & 50.54\% & 88.00\% & 64.20\% & 739 & 103 \\
& SAP-DT & 63.01\% & 58.33\% & 33.45\% & 42.52\% & 205 & 571 \\
& SAP-RF & 62.44\% & 60.17\% & 24.13\% & 34.44\% & 137 & 651 \\
& SAP-XGB & 71.97\% & 73.36\% & 49.42\% & 59.05\% & 154 & 434 \\
& \wb{Cerebro} & \wb{77.98\%} & \wb{83.22\%} & \wb{57.81\%} & \wb{68.23\%} & \wb{100} & \wb{362} \\
& \wb{Hercule} & \wb{83.03\%} & \wb{71.08\%} & \wb{81.35\%} & \wb{75.87\%} & \wb{284} & \wb{160} \\
& SocketAI & 84.80\% & 81.89\% & 80.65\% & 81.27\% & 153 & 166 \\
& \textbf{\tool{} } & \textbf{97.37\%} & \textbf{99.37\%} & \textbf{94.14\%} & \textbf{96.68\%} & \textbf{5} & \textbf{49} \\
\midrule
\multirow{8}{*}{\makecell[l]{Obfuscation \\ Dataset}}
& Bandit4Mal & 54.84\% & 54.31\% & 64.75\% & 59.07\% & 567 & 367 \\
& GuardDog & 89.60\% & 96.96\% & 81.49\% & 88.56\% & 25 & 181 \\
& OSSGadget & 50.95\% & 50.89\% & 85.37\% & 63.77\% & 856 & 152 \\
& PyPI WareCheck & 63.51\% & 58.25\% & 97.02\% & 72.79\% & 724 & 31 \\
& SAP-DT & 52.88\% & 56.16\% & 28.91\% & 38.17\% & 235 & 740 \\
& SAP-RF & 65.83\% & 76.01\% & 46.88\% & 57.99\% & 154 & 553 \\
& SAP-XGB & 60.51\% & 70.44\% & 37.08\% & 48.58\% & 162 & 655 \\
& \wb{Cerebro} & \wb{78.00\%} & \wb{78.93\%} & \wb{76.73\%} & \wb{77.82\%} & \wb{213} & \wb{242} \\
& \wb{Hercule} & \wb{84.34\%} & \wb{85.64\%} & \wb{83.41\%} & \wb{84.51\%} & \wb{145} & \wb{172} \\
& SocketAI & 84.10\% & 82.13\% & 87.42\% & 84.69\% & 198 & 131 \\
& \textbf{\tool{} } & \textbf{98.28\%} & \textbf{97.58\%} & \textbf{99.02\%} & \textbf{98.29\%} & \textbf{25} & \textbf{10} \\
\bottomrule
\end{tabular}
\end{table}
}

\subsection{Dataset}
\label{subsec:eva_dataset}

We constructed four evaluation datasets to assess our knowledge-driven detection framework against different operational challenges, as detailed in Table~\ref{tab:eval_datasets}. The Original Test Dataset comprises our reserved 20\% holdout set with 2,394 benign and 2,363 malicious packages. The Latest Packages Dataset contains 1,001 benign packages randomly sampled from recent PyPI uploads and 1,097 malicious packages from the Open Source Vulnerability Database (OSV)~\cite{osv2024}, collected through July 2024, \wb{which is after the knowledge cutoff date of June 1, 2024 for both GPT-4.1 and GPT-4.1-mini used in our evaluation, ensuring no data leakage in our LLM-based experiments.} The Obfuscated Dataset \wb{was created using} Intensio Obfuscator~\cite{intensio2024} to transform 1,019 benign and 1,050 malicious packages from our test set, evaluating detection resilience when syntactic signatures are deliberately obscured through variable renaming, control flow flattening, and string encoding. 
\wb{The NPM Dataset includes 809 malicious JavaScript packages from Ohm et al.~\cite{ohm2020backstabber} and 953 benign packages randomly sampled from highly downloaded NPM packages}, testing whether behavioral knowledge extracted from Python packages transfers effectively to different programming ecosystems.

\subsection{Baseline Selection}

As shown in Table~\ref{tab:baseline_selection}, various tools have been developed for detecting malicious PyPI packages using rule-based static analysis, dynamic analysis, and ML-based approaches. While numerous tools exist, many lack public availability or source code access, preventing reproducible evaluation. Therefore, we selected baseline tools for comparison based on two criteria: open-source implementation and active maintenance.

Based on these criteria, we selected eight baseline tools for evaluation. Bandit4Mal~\cite{bandit4mal2024}, OSSGadget~\cite{ossgadget2024}, and GuardDog~\cite{guarddog2022} employ rule-based static analysis with predefined patterns for suspicious operations. Bandit4Mal flags security-sensitive APIs, OSSGadget uses regular expressions for pattern matching, and GuardDog implements heuristics for detection. \wb{Hercule~\cite{shariffdeen2025detecting} performs static analysis using CodeQL to detect malicious behaviors.} PyPI WareCheck~\cite{warehouse2020malware} represents PyPI's official malware checks. Machine learning approaches include SAP, which applies Random Forest~(SAP-RF), Decision Tree~(SAP-DT), and XGBoost~(SAP-XGB) classifiers to features extracted from package metadata and code structure. \wb{Cerebro~\cite{zhang2025killing} uses behavior abstraction and BERT models for cross-ecosystem detection on both NPM and PyPI packages.} SocketAI~\cite{zahan2025leveraging} is a purely LLM-based detection method that employs two models in cooperation. In our configuration, we use GPT-4.1 and GPT-4.1-nano for detection. Unlike other tools that output package-level results, SocketAI examines individual Python files and marks the entire package as malicious if any single file is flagged as malicious. These tools provide comprehensive coverage of current detection methodologies, enabling systematic comparison with our knowledge-driven approach.

\subsection{RQ2: Accuracy}

Table~\ref{tab:rq2_results} presents detection performance across three evaluation datasets. \tool{} analyzes individual Python files within each package, flagging a package as malicious if any file is detected as malicious. \wb{\tool{} achieves 99.50\% accuracy on the original dataset, substantially outperforming GuardDog's 94.12\% accuracy. Bandit4Mal achieved only 33.37\% accuracy with 1,927 false positives and 1,080 false negatives.}


On the evaluation dataset, traditional tools exhibit context-blind detection that generates excessive false positives. GuardDog achieves the lowest baseline false positive rate at 4.30\%, yet 58.02\% of its errors stem from code-execution flags on legitimate operations such as \texttt{exec(compile(open('version.py').read(), 'version.py', 'exec'))} used for version checking. Bandit4Mal demonstrates a catastrophic 71.01\% false positive rate, with 1,682 files (87.97\%) flagged for url\_found alerts that include project homepages and documentation links, alongside 56.28\% and 39.75\% false positive rates for basic read and write operations, respectively. The tool's overly broad rules generate an average of 5.37 detection types per file, exemplified by \texttt{os\_sys-2.1.4}, which triggers 48 distinct alerts. PyPI WareCheck employs coarse-grained YARA rules that flag 97.00\% of packages for metaprogramming\_in\_setup merely for utilizing standard Python imports, with one file generating 31,242 individual alerts. OSSGadget's 83.23\% false positive rate results from broad LOLBAS~\cite{fortinet2025} rules that flag 96.97\% of packages for Linux commands, 87.13\% for Windows utilities, and 71.85\% for data exfiltration when packages execute routine file operations. \wb{In contrast, \tool{} achieves only 2 false positives and 19 false negatives. Among the 19 false negatives, 8 packages involve dependency confusion where the analyzed package does not contain malicious code but imports a malicious dependency, 6 packages install code from remote repositories at runtime, and 5 packages embed malicious payloads in non-Python files. These cases fall outside \tool{}'s design scope of Python code behavioral analysis.}

\wb{Hercule employs CodeQL-driven static analysis to detect five predefined behavior categories: exfiltration, file operations, network activities, obfuscation, and process manipulation. While it achieves 88.76\% accuracy on the evaluation dataset, its analysis can take a long time for complex packages. We set a one-hour timeout per package, after which packages are classified as benign by default. Only a small number of packages exceeded this timeout: 112 on evaluation dataset (109 benign, 3 malicious), 10 on latest dataset (all benign), and none on obfuscation dataset. Since timed-out benign packages are correctly classified by default, the timeout has minimal impact on the reported metrics. Its rigid rule definitions also limit adaptability. When applied to newer packages, accuracy drops to 83.03\% as novel attack patterns fall outside its predefined rules. Cerebro constructs behavior sequences and fine-tunes a BERT model as a binary classifier. It achieves 89.12\% accuracy on the evaluation dataset but suffers significant degradation on obfuscated code (78.00\%), as syntactic transformations disrupt the token-level features that BERT relies on for classification.}

The latest package dataset reveals adaptation limitations in rule-based systems. GuardDog's recall drops from 94.42\% to 68.12\% when encountering newer packages, with false positives shifting to concentrate on shady-links detection rather than code-execution patterns. \tool{} maintains 94.14\% recall with only 49 false negatives, demonstrating resilience to evolving attack techniques. Obfuscation resistance varies dramatically across detection approaches. Machine learning models (SAP-DT, SAP-RF, SAP-XGB) exhibit 20-30\% accuracy degradation when code transformations alter feature distributions. SAP-RF's recall drops from 78.05\% to 46.88\%, missing 553 malicious packages after obfuscation. Rule-based tools show mixed results: GuardDog maintains 89.60\% accuracy through semantic-aware heuristics, while Bandit4Mal and OSSGadget hover around 50\% accuracy. Our framework achieves 98.28\% accuracy on obfuscated packages by analyzing behavioral sequences rather than syntactic patterns since variable renaming and control flow obfuscation do not alter the underlying API execution order that our pattern mining captures.

SocketAI, as an LLM-based approach, achieves 90.95\% accuracy on the evaluation dataset, which is the highest among traditional baselines. Yet, it still generates 161 false positives and demonstrates performance degradation on newer packages (84.80\%) and obfuscated code (84.10\%). This decline reveals the limitations of relying solely on LLM general knowledge without domain-specific behavioral understanding. In contrast, our approach systematically extracts and summarizes knowledge from both malicious and false positive detection patterns, creating a specialized knowledge base that enables more precise contextual reasoning and maintains robust performance across all evaluation scenarios.

The performance differential stems from architectural differences in detection logic. Traditional tools scan code statements or extract statistical features, losing behavioral context during analysis. Our approach preserves execution sequences: a reverse shell pattern remains detectable as $[create\_socket, establish\_tcp\_connection, dup\_socket\_stdin \\, dup\_socket\_stdout, dup\_socket\_stderr]$ regardless of implementation variations. \wb{This sequence is inherently malicious because it initiates an outbound connection to a remote server and redirects standard I/O streams (stdin, stdout, stderr) to the socket, enabling attackers to interactively execute commands on the compromised machine.} On our evaluation dataset, deterministic pattern matching identified 1,763 malicious packages and flagged 873 benign packages with justifiable patterns. The knowledge-driven LLM successfully filtered these false positives through contextual reasoning, achieving detection accuracy that exceeds the best baseline by 5.38\% while reducing false positives by 98.58\%.

\responsebox{Response to RQ2: Our knowledge-driven method achieves 99.50\% accuracy with only 2 false positives compared to baselines. The behavioral knowledge enables consistent performance across emerging threats (97.37\%) and obfuscated code (98.28\%), where traditional tools drop to about 50\% accuracy due to syntactic pattern disappearance.}

\subsection{RQ3: Ablation}

\begin{table*}[htbp]
\centering
\caption{Knowledge Enhancement on Malicious Package Detection against Existing Tools}
\label{tab:rq3_ablation}
\scriptsize
\begin{tabular}{llrrrrrr}
\toprule
\textbf{Dataset} & \textbf{Method} & \textbf{Accuracy} & \textbf{Precision} & \textbf{Recall} & \textbf{F1-Score} & \textbf{FP} & \textbf{FN} \\
\midrule
\multirow{4}{*}{\parbox{1.5cm}{\centering Evaluation\\Dataset}} 
& OSSGadget (Only) & 50.01\% & 48.13\% & 88.38\% & 62.32\% & 2,000 & 244 \\
& OSSGadget + Knowledge & {\color{red}$\uparrow$\tiny\textit{39.90\%}} 89.91\% & {\color{red}$\uparrow$\tiny\textit{42.42\%}} 90.55\% & {\color{orange}$\downarrow$\tiny\textit{0.81\%}} 87.57\% & {\color{red}$\uparrow$\tiny\textit{26.71\%}} 89.03\% & {\color{red}$\uparrow$\tiny\textit{1808}} 192 & {\color{orange}$\downarrow$\tiny\textit{17}} 261 \\
& Bandit4Mal (Only) & 33.37\% & 34.72\% & 48.69\% & 40.54\% & 1,927 & 1,080 \\
& Bandit4Mal + Knowledge & {\color{red}$\uparrow$\tiny\textit{37.16\%}} 70.53\% & {\color{red}$\uparrow$\tiny\textit{46.56\%}} 81.28\% & {\color{orange}$\downarrow$\tiny\textit{0.85\%}} 47.84\% & {\color{red}$\uparrow$\tiny\textit{19.69\%}} 60.23\% & {\color{red}$\uparrow$\tiny\textit{1695}} 232 & {\color{orange}$\downarrow$\tiny\textit{18}} 1,098 \\
\midrule
\multirow{4}{*}{\parbox{1.5cm}{\centering Latest\\Dataset}} 
& OSSGadget (Only) & 52.63\% & 45.79\% & 86.33\% & 59.84\% & 875 & 117 \\
& OSSGadget + Knowledge & {\color{red}$\uparrow$\tiny\textit{38.82\%}} 91.45\% & {\color{red}$\uparrow$\tiny\textit{51.16\%}} 96.95\% & {\color{orange}$\downarrow$\tiny\textit{4.67\%}} 81.66\% & {\color{red}$\uparrow$\tiny\textit{28.81\%}} 88.65\% & {\color{red}$\uparrow$\tiny\textit{853}} 22 & {\color{orange}$\downarrow$\tiny\textit{40}} 157 \\
& Bandit4Mal (Only) & 58.45\% & 49.60\% & 94.40\% & 65.03\% & 822 & 48 \\
& Bandit4Mal + Knowledge & {\color{red}$\uparrow$\tiny\textit{36.01\%}} 94.46\% & {\color{red}$\uparrow$\tiny\textit{47.24\%}} 96.84\% & {\color{orange}$\downarrow$\tiny\textit{5.02\%}} 89.38\% & {\color{red}$\uparrow$\tiny\textit{27.93\%}} 92.96\% & {\color{red}$\uparrow$\tiny\textit{797}} 25 & {\color{orange}$\downarrow$\tiny\textit{43}} 91 \\
\midrule
\multirow{4}{*}{\parbox{1.5cm}{\centering Obfuscation\\Dataset}} 
& OSSGadget (Only) & 50.95\% & 50.89\% & 85.37\% & 63.77\% & 856 & 152 \\
& OSSGadget + Knowledge & {\color{red}$\uparrow$\tiny\textit{39.56\%}} 90.51\% & {\color{red}$\uparrow$\tiny\textit{45.48\%}} 96.37\% & {\color{orange}$\downarrow$\tiny\textit{0.96\%}} 84.41\% & {\color{red}$\uparrow$\tiny\textit{26.22\%}} 89.99\% & {\color{red}$\uparrow$\tiny\textit{823}} 33 & {\color{orange}$\downarrow$\tiny\textit{10}} 162 \\
& Bandit4Mal (Only) & 54.84\% & 54.31\% & 64.75\% & 59.07\% & 567 & 367 \\
& Bandit4Mal + Knowledge & {\color{red}$\uparrow$\tiny\textit{25.48\%}} 80.32\% & {\color{red}$\uparrow$\tiny\textit{41.24\%}} 95.55\% & {\color{orange}$\downarrow$\tiny\textit{0.87\%}} 63.88\% & {\color{red}$\uparrow$\tiny\textit{16.50\%}} 76.57\% & {\color{red}$\uparrow$\tiny\textit{536}} 31 & {\color{orange}$\downarrow$\tiny\textit{9}} 376 \\
\bottomrule
\end{tabular}
\end{table*}

\begin{table*}[htbp]
\centering
\caption{Knowledge Enhancement on Malicious Package Detection against LLMs}
\label{tab:ablation_study}
\scriptsize
\setlength{\tabcolsep}{3pt}
\begin{tabular}{llrrrrrr}
\toprule
\textbf{Dataset} & \textbf{Method} & \textbf{Accuracy} & \textbf{Precision} & \textbf{Recall} & \textbf{F1-Score} & \textbf{FP} & \textbf{FN} \\
\midrule
\multirow{3}{*}{\parbox{1.5cm}{\centering Evaluation\\Dataset}} 
& GPT-4.1 & {\color{orange}$\downarrow$\tiny\textit{2.78\%}} 96.72\% & {\color{orange}$\downarrow$\tiny\textit{5.16\%}} 94.74\% & {\color{orange}$\downarrow$\tiny\textit{0.70\%}} 98.38\% & {\color{orange}$\downarrow$\tiny\textit{2.96\%}} 96.53\% & {\color{orange}$\downarrow$\tiny\textit{113}} 115 & {\color{orange}$\downarrow$\tiny\textit{15}} 34 \\
& GPT-4.1-Mini + Knowledge & {\color{red}$\uparrow$\tiny\textit{0.01\%}} 99.51\% & {\color{orange}$\downarrow$\tiny\textit{0.33\%}} 99.57\% & {\color{red}$\uparrow$\tiny\textit{0.29\%}} 99.37\% & {\color{orange}$\downarrow$\tiny\textit{0.02\%}} 99.47\% & {\color{orange}$\downarrow$\tiny\textit{7}} 9 & {\color{red}$\uparrow$\tiny\textit{6}} 13 \\
& \textbf{\tool{} (GPT-4.1 + Knowledge)} & \textbf{99.50\%} & \textbf{99.90\%} & \textbf{99.08\%} & \textbf{99.49\%} & \textbf{2} & \textbf{19} \\
\midrule
\multirow{3}{*}{\parbox{1.5cm}{\centering Latest\\Dataset}} 
& GPT-4.1 & {\color{orange}$\downarrow$\tiny\textit{9.91\%}} 87.46\% & {\color{orange}$\downarrow$\tiny\textit{2.73\%}} 96.64\% & {\color{orange}$\downarrow$\tiny\textit{15.38\%}} 78.76\% & {\color{orange}$\downarrow$\tiny\textit{9.89\%}} 86.79\% & {\color{orange}$\downarrow$\tiny\textit{25}} 30 & {\color{orange}$\downarrow$\tiny\textit{184}} 233 \\
& GPT-4.1-Mini + Knowledge & {\color{red}$\uparrow$\tiny\textit{0.23\%}} 97.60\% & {\color{orange}$\downarrow$\tiny\textit{2.22\%}} 97.15\% & {\color{red}$\uparrow$\tiny\textit{2.78\%}} 96.92\% & {\color{red}$\uparrow$\tiny\textit{0.35\%}} 97.03\% & {\color{orange}$\downarrow$\tiny\textit{19}} 24 & {\color{orange}$\downarrow$\tiny\textit{23}} 26 \\
& \textbf{\tool{} (GPT-4.1 + Knowledge)} & \textbf{97.37\%} & \textbf{99.37\%} & \textbf{94.14\%} & \textbf{96.68\%} & \textbf{5} & \textbf{49} \\
\midrule
\multirow{3}{*}{\parbox{1.5cm}{\centering Obfuscation\\Dataset}} 
& GPT-4.1 & {\color{orange}$\downarrow$\tiny\textit{29.70\%}} 68.58\% & {\color{orange}$\downarrow$\tiny\textit{35.84\%}} 61.74\% & {\color{orange}$\downarrow$\tiny\textit{0.27\%}} 98.75\% & {\color{orange}$\downarrow$\tiny\textit{22.31\%}} 75.98\% & {\color{orange}$\downarrow$\tiny\textit{612}} 637 & {\color{orange}$\downarrow$\tiny\textit{3}} 13 \\
& GPT-4.1-Mini + Knowledge & {\color{orange}$\downarrow$\tiny\textit{0.63\%}} 97.65\% & {\color{orange}$\downarrow$\tiny\textit{1.03\%}} 96.55\% & {\color{orange}$\downarrow$\tiny\textit{0.20\%}} 98.82\% & {\color{orange}$\downarrow$\tiny\textit{0.62\%}} 97.67\% & {\color{orange}$\downarrow$\tiny\textit{11}} 36 & {\color{orange}$\downarrow$\tiny\textit{2}} 12 \\
& \textbf{\tool{} (GPT-4.1 + Knowledge)} & \textbf{98.28\%} & \textbf{97.58\%} & \textbf{99.02\%} & \textbf{98.29\%} & \textbf{25} & \textbf{10} \\
\bottomrule
\end{tabular}
\footnotesize
\begin{enumerate}
    \item \tool{} (GPT-4.1 + Knowledge): Full GPT-4.1 model enhanced with our Knowledge framework.
\end{enumerate}
\end{table*}

\begin{table}[t]
\centering
\caption{\wb{Performance with Different LLM Backends on the Latest Dataset}}
\label{tab:opensource_llms}
\footnotesize
\setlength{\tabcolsep}{4pt}
\wb{
\begin{tabular}{lrrrr}
\toprule
\textbf{Method} & \textbf{Accuracy} & \textbf{Precision} & \textbf{Recall} & \textbf{F1-Score} \\
\midrule
\tool{} (GPT-4.1) & 97.37\% & 99.37\% & 94.14\% & 96.68\% \\
\tool{} (Qwen3-8B) & 95.24\% & 96.86\% & 92.55\% & 94.65\% \\
\tool{} (DeepSeek-V3) & \textbf{97.74\%} & \textbf{98.30\%} & \textbf{96.77\%} & \textbf{97.53\%} \\
\bottomrule
\end{tabular}
}
\end{table}

We designed comparative experiments to evaluate RAG knowledge enhancement effects. Table~\ref{tab:rq3_ablation} compares original detection results from static analysis tools with performance after RAG framework integration. The experimental workflow: static tools (OSSGadget, Bandit4Mal) detect and mark suspicious locations, then we extract corresponding code snippets and utilize the RAG framework to retrieve relevant behavioral patterns for secondary analysis and reclassification. Table~\ref{tab:ablation_study} compares different LLM configurations: GPT-4.1 alone relies solely on pre-trained knowledge for direct code analysis, GPT-4.1-Mini+Knowledge combines a lightweight model with our extracted pattern and sequence knowledge, and GPT-4.1+Knowledge represents the complete solution. This design isolates and quantifies the independent contribution of RAG knowledge enhancement.

Knowledge enhancement dramatically improves static analysis tool performance across different limitation profiles. OSSGadget's accuracy increases from 50.01\% to 89.91\% while false positives drop from 2,000 to 192. The tool maintains high recall (88.38\%) but generates excessive false positives through overly broad pattern matching. Adding behavioral knowledge preserves recall at 87.57\% while increasing precision from 48.13\% to 90.55\%. Bandit4Mal exhibits the opposite problem: low recall (48.69\%) \wb{due to} conservative rule definitions. RAG enhancement maintains similar recall but increases precision from 34.72\% to 81.28\% and overall accuracy from 33.37\% to 70.53\%. These improvements stem from behavioral patterns providing semantic context that distinguishes legitimate API usage from malicious intent based on execution sequences rather than individual function calls.

Language model comparisons reveal that knowledge quality trumps model capacity. GPT-4.1 alone achieves 96.72\% accuracy on normal packages but degrades to 68.58\% on obfuscated code, generating 637 false positives when syntactic cues disappear. GPT-4.1-Mini with Knowledge reaches 99.51\% accuracy, which surpasses the larger model without Knowledge by 2.79\%, demonstrating that 304 behavioral patterns provide detection signals that raw language understanding cannot extract. This obfuscation resistance extends to enhanced static tools: OSSGadget+Knowledge maintains 90.51\% accuracy on obfuscated code versus 50.95\% baseline performance. The improvement stems from sequence-level patterns that persist through variable renaming and control flow transformations, remaining detectable regardless of syntactic obfuscation.

\wb{To evaluate reproducibility and reduce dependence on closed-source models, we tested \tool{} with open-source LLMs on the Latest Dataset. As shown in Table~\ref{tab:opensource_llms}, DeepSeek-V3 achieves 97.74\% accuracy, slightly outperforming GPT-4.1 (97.37\%). Qwen3-8B achieves 95.24\% accuracy with acceptable performance degradation. These results demonstrate that \tool{} generalizes across different LLM backends and does not require proprietary APIs.}

\responsebox{Response to RQ3: RAG knowledge enhancement increases OSSGadget's accuracy by 39.90\% and Bandit4Mal's by 37.16\%, while reducing their false positives by 90.4\% and 88.0\% respectively. The behavioral knowledge \wb{enables} GPT-4.1-Mini to achieve 99.51\% accuracy, demonstrating that extracted sequences provide critical detection signals independent of model capacity.}

\subsection{RQ4: Usability}
\label{subsec:usability}

\begin{figure}[t]
    \centering
    \includegraphics[width=0.5\textwidth]{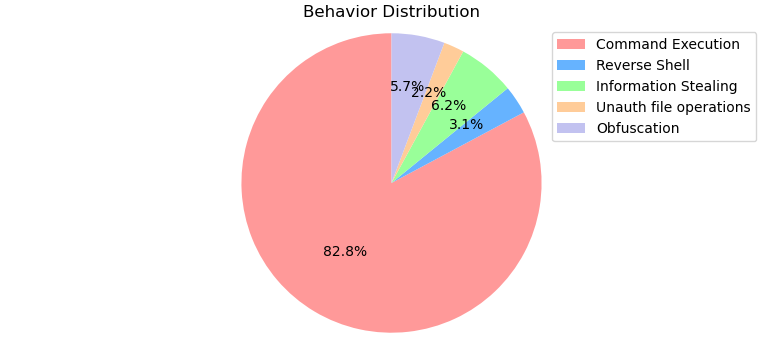}
    \caption{\wb{Distribution of Malicious Package Behaviors}}
    \label{fig:behavior_distribution}
\end{figure}

To evaluate \tool{}'s real-world effectiveness, we deployed \tool{} on \textit{PyPI.org} from March to June 2024. During this three-month period, 8,249 new packages were released on \textit{PyPI.org}. Our system detected 233 packages as malicious, identifying 219 previously unknown malicious packages encompassing 287 versions that were not listed in the \textit{OSV} or \textit{Snyk} databases.
\wb{Following responsible disclosure practices, each detected package underwent manual verification by our research team before submission to the PyPI security team via their official platform (\url{https://pypi.org/security/}).} \wb{All 219 reported malicious packages were confirmed and subsequently removed by official PyPI maintainers, with 253 official acknowledgment letters received}. \wb{We did not disclose any package information publicly before removal. We notified downstream mirror maintainers only after PyPI had confirmed and removed the packages.} In contrast, baseline tools generated over 7,342 false positives on legitimate packages during the same period, highlighting the practical superiority of our approach in production environments. Table \ref{table:new_packages} shows a subset of the new malicious packages we detected. By analyzing Google Cloud's PyPI package download data, we found that these malicious packages were downloaded 39,420 times in three months. Approximately 37,584 downloads (95.1\%) utilized the \textit{sdist} (source code distribution) method, while the remaining downloads were conducted via \textit{bdist\_wheel} (binary distribution). Regarding geographical distribution, we observed that 66.5\% of these downloads originated from the United States. Furthermore, China, Germany, and Singapore accounted for 8.4\%, 7.4\%, and 3.2\% of the downloads, respectively. Source code analysis showed that 94.4\% of malicious packages used \textbf{\textit{install-time}} attacks. These attacks embed malicious code in the \textit{setup.py} file or overwrite \textit{CustomInstallCommand} class function, and trigger when the package is installed. Another 4.2\% of malicious packages employed the \textbf{\textit{import-time}} attack, inserting malicious code into \textit{\_\_init\_\_.py} file, which activates when a user imports the package. The remaining malicious packages utilized the \textbf{\textit{run-time}} attack method. \wb{Figure~\ref{fig:behavior_distribution} shows the distribution of malicious behaviors in the newly detected packages.} \textit{Command execution} was the most prominent malicious behavior, occurring in 188 packages, accounting for 82.8\% of the total. Additionally, information theft and code obfuscation were observed in 14 and 13 packages, respectively. Reverse shell behavior had a lower incidence, appearing in 7 packages, making up 3.1\%. In the package metadata analysis, the majority of malicious packages employed \wb{typosquatting attacks}, misleading developers by mimicking the names of legitimate packages.

\begin{table*}[t]
\centering
\caption{Some New Malicious Packages Detected from PyPI.org by \tool{}}
\small
\renewcommand{\arraystretch}{1.0}
\setlength{\tabcolsep}{2pt}
\begin{adjustbox}{max width=\textwidth}
\begin{tabular}{|c|c|c|c|c|c|c|c|c|}
\hline
\textbf{Package Name} & \textbf{   Versions  } & \textbf{  Location  } & \textbf{Command Execution} & \textbf{Information Stealing} & \textbf{Reverse Shell} & \textbf{ Obfuscation } & \textbf{ File Operation } & \textbf{ Downloads } \\
\hline
bussardweg4av3 & 1.0.0 & setup.py & \checkmark & - & - & - & - & 106 \\
\hline
pytyon & 1.0.0 & setup.py & \checkmark & - & - & - & - & 102 \\
\hline
eutherium & 1.0.0 & setup.py & \checkmark & - & - & - & - & 199 \\
\hline
openeaa & 1.0.0 & setup.py & \checkmark & - & - & - & - & 245 \\
\hline
class-py & 1.0.0 & setup.py & - & - & \checkmark & - & - & 107 \\
\hline
quickwebbasicauth & 2.3.2 & setup.py & - & - & \checkmark & - & \checkmark & 307 \\
\hline
artifact-lab-3-package-e90915e1 & 0.1.1, 0.1.3, 0.1.4 & setup.py & \checkmark & - & - & - & \checkmark & 466 \\
\hline
testkaralpoc45654 & 1.0.0 & setup.py & - & \checkmark & - & - & - & 176 \\
\hline
google-requests & 99.3.9 & pre\_install.py & \checkmark & - & - & - & - & 164 \\
\hline
lyft-service & 9.99.1 & pre\_install.py & \checkmark & - & - & - & - & 205 \\
\hline
jupyter-calendar-extension & 0.1 & pre\_install.py & \checkmark & - & - & - & - & 88 \\
\hline
networkx-match-algr & 0.1.1 & core.py & - & - & - & - & - & 353 \\
\hline
pyjous & 1.0.2 & setup.py & \checkmark & - & - & \checkmark & - & 113 \\
\hline
pyzelf & 2.0.1 & setup.py & \checkmark & - & - & \checkmark & - & 123 \\
\hline
thesis-package & 1.0.0 & setup.py & - & - & - & - & \checkmark & 84 \\
\hline
thesis-uniud-package & 1.0.0 & setup.py & - & - & - & - & \checkmark & 118 \\
\hline
builderknower2 & 0.1.12 \textasciitilde 0.1.30 & setup.py & \checkmark & - & - & - & - & 3,137 \\
\hline
booto3 & 0.0.1 & setup.py & \checkmark & - & - & - & - & 108 \\
\hline
utilitytools & 0.0.2 \textasciitilde 0.0.9 & \_\_init\_\_.py & - & - & - & - & \checkmark & 1,152 \\
\hline
utilitytool & 0.0.2 & \_\_init\_\_.py & - & - & - & - & \checkmark & 95 \\
\hline
nt4padyp3 & 0.0.2 & setup.py & \checkmark & - & - & - & - & 340 \\
\hline
importlib-metadate & 99.99 & setup.py & \checkmark & - & - & - & - & 229 \\
\hline
reqwestss & 0.1.0 & index.py & - & - & \checkmark & - & - & 242 \\
\hline
numberpy & 0.1.0 & index.py & - & - & \checkmark & - & - & 249 \\
\hline
defca & 3.0.0 & test.py & \checkmark & - & - & \checkmark & - & 444 \\
\hline
builderknower & 0.1.8 \textasciitilde 0.1.12 & setup.py & \checkmark & - & - & - & - & 1,409 \\
\hline
rev0001q1 & 2.0.0 & setup.py & - & - & \checkmark & - & - & 268 \\
\hline
revabc01q1 & 0.0.2 & setup.py & - & - & \checkmark & - & - & 196 \\
\hline
\end{tabular}
\end{adjustbox}
\label{table:new_packages}
\footnotesize
\begin{enumerate}
    \item The symbol \checkmark indicates the presence of such information, while the symbol '-' indicates its absence. Location: the file where the malicious code is located. Downloads: the number of times the malicious package has been downloaded.
\end{enumerate}
\end{table*}

\begin{table}[t]
   \centering
   \caption{Number of Malicious Packages in Mirrors}
   \label{table:mirror-sources}
   \footnotesize
   \begin{tabular}{lccc}
       \toprule
       \textbf{Mirror Source} & \textbf{Package Nums} & \textbf{Version Nums} \\
       \midrule
       Tencent Mirror & 1,225 & 1,650 \\
       Huawei Mirror & 2 & 5 \\
       Douban Mirror & 1,219 & 1,649 \\
       Aliyun Mirror & 1 & 3 \\
       Tsinghua Mirror & 1,039 & 2,010 \\
       BFSU Mirror & 1,034 & 2,005 \\
       \bottomrule
   \end{tabular}
\end{table}

Simultaneously, we conducted an analysis of the major downstream mirrors of PyPI and identified over 4,500 malicious packages across 7,200 versions. Among these, the Tsinghua, Tencent, Douban, and BFSU mirrors contained the highest number of malicious packages, each exceeding 1,000, as shown in Table \ref{table:mirror-sources}. Notably, the Tsinghua mirror had more than 2,010 affected versions. In contrast, the Aliyun and Huawei mirrors were relatively secure, with only a small number of malicious packages detected. These residual malicious packages have been downloaded more than 200,000 times, posing a significant threat to user security. Using \tool{}, we detected and identified these malicious packages and informed the maintainers of the downstream mirrors. Both Tsinghua and Tencent have confirmed the issue and have removed the malicious packages.

\wb{\tool{} is fully automated. Package files without sensitive APIs need no analysis. Files with sensitive APIs require 1-3 LLM queries depending on pattern matching results: context extraction, action mapping, and RAG-based analysis if deterministic patterns are not matched. On average, \tool{} takes 34.09 seconds per package (8.8 seconds for malicious packages, while benign packages are usually larger and require more time). This is slower than static analysis tools such as Bandit4Mal (1.50s), OSSGadget (4.13s), and GuardDog (25.10s), but significantly faster than LLM-based SocketAI (164.67s) and \wb{the dynamic analysis tool Hercule} (578.69s). The efficiency gain over SocketAI comes from our hybrid strategy where RAG-based analysis runs only when deterministic patterns are not matched. Regarding cost, \tool{} requires \$0.023 per package compared to SocketAI's \$0.156. When using open-source models such as DeepSeek-V3, the inference cost can be eliminated entirely.}

\responsebox{Response to RQ4: \tool{} demonstrated strong practical utility in real-world application, successfully identifying 219 newly discovered malicious packages (confirmed by PyPI officials) and over 4,500 malicious packages in major downstream mirrors.}

\subsection{RQ5: Cross-ecosystem}

\begin{table*}[htbp]
\centering
\caption{Cross-language applicability: Performance evaluation on different programming languages}
\footnotesize
\label{tab:language_applicability}
\setlength{\tabcolsep}{4pt}
\begin{tabular}{lrrrrrr}
\toprule
\textbf{Method} & \textbf{Accuracy} & \textbf{Precision} & \textbf{Recall} & \textbf{F1-Score} & \textbf{FP} & \textbf{FN} \\
\midrule
GuardDog & {\color{orange}$\downarrow$\tiny\textit{3.97\%}} 94.10\% & {\color{orange}$\downarrow$\tiny\textit{0.82\%}} 97.44\% & {\color{orange}$\downarrow$\tiny\textit{8.04\%}} 89.49\% & {\color{orange}$\downarrow$\tiny\textit{4.59\%}} 93.30\% & {\color{orange}$\downarrow$\tiny\textit{5}} 19 & {\color{orange}$\downarrow$\tiny\textit{1}} 85 \\
OSSGadget & {\color{orange}$\downarrow$\tiny\textit{44.27\%}} 53.80\% & {\color{orange}$\downarrow$\tiny\textit{48.43\%}} 49.83\% & {\color{orange}$\downarrow$\tiny\textit{7.42\%}} 90.11\% & {\color{orange}$\downarrow$\tiny\textit{33.72\%}} 64.17\% & {\color{orange}$\downarrow$\tiny\textit{41}} 734 & {\color{orange}$\downarrow$\tiny\textit{1}} 80 \\
\wb{Cerebro} & \wb{{\color{orange}$\downarrow$\tiny\textit{10.75\%}} 87.32\%} & \wb{{\color{green}$\uparrow$\tiny\textit{1.40\%}} 99.66\%} & \wb{{\color{orange}$\downarrow$\tiny\textit{25.42\%}} 72.11\%} & \wb{{\color{orange}$\downarrow$\tiny\textit{14.21\%}} 83.68\%} & \wb{{\color{green}$\uparrow$\tiny\textit{12}} 2} & \wb{{\color{orange}$\downarrow$\tiny\textit{206}} 226} \\
GPT-4.1 & {\color{orange}$\downarrow$\tiny\textit{1.36\%}} 96.71\% & {\color{orange}$\downarrow$\tiny\textit{0.18\%}} 98.08\% & {\color{orange}$\downarrow$\tiny\textit{2.85\%}} 94.68\% & {\color{orange}$\downarrow$\tiny\textit{1.54\%}} 96.35\% & {\color{orange}$\downarrow$\tiny\textit{1}} 15 & {\color{orange}$\downarrow$\tiny\textit{23}} 43 \\
\textbf{\tool{} (RAG + GPT-4.1)} & \textbf{98.07\%} & \textbf{98.26\%} & \textbf{97.53\%} & \textbf{97.89\%} & \textbf{14} & \textbf{20} \\
\bottomrule
\end{tabular}
\end{table*}

To evaluate cross-ecosystem generalization, we assessed our approach on 1,762 NPM packages (953 benign, 809 malicious) using established baselines for JavaScript malware detection. We compared against GuardDog and OSSGadget, the most widely adopted static analysis tools for NPM malicious package detection, \wb{Cerebro~\cite{zhang2025killing}, which performs cross-platform detection using behavior abstraction and BERT models,} and GPT-4.1 as a foundational baseline. GuardDog and OSSGadget apply their native rule-based detection directly to NPM packages, while both GPT-4.1 and our knowledge-driven method analyze individual JavaScript files within each package. A package receives a malicious classification if any constituent file is flagged as malicious. \wb{We directly apply PyPI-derived patterns to NPM packages without any adaptation or ecosystem-specific modifications. The detection workflow remains identical to PyPI detection.}

Table~\ref{tab:language_applicability} demonstrates that \tool{} achieves superior cross-language performance with 98.07\% accuracy, substantially outperforming all baselines. \tool{} significantly outperforms rule-based approaches, achieving 3.97\% higher accuracy than GuardDog and 44.27\% higher than OSSGadget. \wb{Compared to Cerebro, \tool{} achieves 10.75\% higher accuracy and 25.42\% higher recall, reducing false negatives from 226 to 20. The GPT-4.1 foundation model reaches 96.71\% accuracy, while our approach delivers a 1.36\% improvement and reduces false negatives from 43 to 20.} The RAG enhancement maintains high precision at 98.26\% and strong recall at 97.53\%, \wb{with 14 false positives compared to Cerebro's 2 and} OSSGadget's 734, demonstrating that behavioral pattern abstractions facilitate effective cross-language transfer beyond pure language model capabilities.

This performance differential stems from fundamental limitations in rule-based detection approaches when applied across ecosystems. GuardDog's JavaScript rules rely on specific syntactic constructs such as \texttt{child\_process.exec} calls and obfuscation patterns like \texttt{while (!![])} loops, but struggle with semantically equivalent malicious behaviors that use different implementation approaches, resulting in 3.97\% lower accuracy compared to \tool{}. OSSGadget's NPM detection depends on literal string matching and basic regular expressions that capture surface-level indicators but lack contextual understanding, achieving 44.27\% lower accuracy than our approach when legitimate packages employ similar constructs for benign purposes. These approaches fundamentally operate on language-specific syntactic signatures that cannot capture underlying behavioral semantics. Conversely, our behavioral abstraction approach captures semantic intent rather than syntactic patterns. Concepts like "process manipulation," "network communication," and "data exfiltration" manifest consistently across programming languages through different but functionally equivalent API calls. The PyPI-derived behavioral knowledge successfully transfers to NPM detection because malicious behaviors follow similar logical patterns regardless of implementation language: establishing network connections, executing system commands, and collecting sensitive information represent universal attack primitives. \wb{\tool{} has higher false positives than Cerebro because NPM has language-specific code semantics like npm script hooks that are not covered by PyPI-derived patterns.} This semantic-level understanding enables effective cross-ecosystem knowledge transfer, demonstrating that behavioral patterns extracted from one package ecosystem can enhance malicious package detection across different programming environments.

\responsebox{Response to RQ5: Our pattern knowledge demonstrates effective cross-ecosystem generalization, achieving 98.07\% accuracy on NPM packages and outperforming GuardDog by 3.97\%, \wb{Cerebro by 10.75\%,} and OSSGadget by 44.27\%. PyPI-derived behavioral knowledge successfully transfers to JavaScript malware detection, demonstrating effective pattern generalization across package ecosystems.}
\section{Limitation and Threats to Validity}
\label{sec:limitation}

\wb{
\noindent\textbf{Knowledge Base Update.} 
\tool{} supports incremental updates. When novel attack techniques emerge, new patterns can be added to the existing knowledge base without rebuilding from scratch. In practice, such updates are rare. Most emerging malware reuses behavioral patterns that our knowledge base already covers.
}

\wb{
\noindent\textbf{Pattern-based Limitations.} 
\tool{} performs code-level detection and cannot identify dependency confusion~\cite{guo2023empirical} attacks where malicious code resides in referenced packages rather than the analyzed package. It also cannot detect threats in remote repository installations or non-Python files embedded in packages. Attackers might try inserting benign sequences to evade detection. However, this evasion strategy is ineffective because our matching operates on minimum subsequences that do not need to appear consecutively. Our behavioral abstraction approach analyzes execution semantics rather than code syntax, which provides resilience against obfuscation techniques.
}

\wb{
\noindent\textbf{LLM Dependency.} 
LLM hallucination is a potential concern. We mitigate this risk through two strategies. First, we use deterministic pattern matching as the primary detection method. Second, when patterns do not match, we retrieve similar cases from the knowledge base as few-shot examples to guide the LLM's reasoning. This grounds the LLM output in real-world examples rather than relying solely on its internal knowledge. We tested \tool{} with both closed-source models (GPT-4.1, GPT-4.1-mini) and open-source models (DeepSeek-V3, Qwen3-8B). All models achieved over 95\% accuracy on the Latest dataset. As LLM capabilities improve, we expect \tool{}'s performance to improve as well.
}

\wb{
\noindent\textbf{Potential Attacks on Detection.} 
Adversaries may attempt to evade or attack \tool{}. Very long contexts may present challenges for LLM-based analysis, but segmented analysis can address this. Prompt injection is another concern. Our slicing approach extracts only relevant code contexts instead of entire files. This reduces the attack surface and limits opportunities for injected prompts to affect detection. Existing LLM defense techniques can also be integrated to further mitigate these risks~\cite{10.1145/3712001, 2024yupeiusenix}.
}
\section{Related Works}
\label{sec:relatedwork}
In this section, we introduce the related works on malicious code detection and other security issues for PyPI packages.

\noindent \textbf{Malicious Code Detection.}
Malicious code detection represents a critical research direction in cybersecurity, with researchers achieving significant progress in static analysis, dynamic monitoring, and deep learning methods. In the Python security domain, PyXhon~\cite{sun2012pyxhon} integrates static analysis and dynamic monitoring techniques to achieve precise function call tracing and security risk assessment. PyComm~\cite{zhou2022pycomm} constructs a machine learning model based on statistical attributes and string sequences. PBDT~\cite{fang2021pbdt} combines call attributes, text statistics, and opcode sequence features for backdoor detection. The application of deep learning methods has further extended the boundaries of detection techniques. MSDT~\cite{tsfaty2022malicious} proposes a static analysis method based on vector space representation to identify injection-type malicious code through anomaly detection strategies. ScriptNet~\cite{stokes2019scriptnet} utilizes deep learning models~\cite{gonzalez2021anomalicious} to process JavaScript files as byte sequences, enabling multi-level detection and classification. Additionally, researchers have developed specialized detection frameworks for specific scripting languages such as PowerShell~\cite{hendler2020amsi} and JavaScript~\cite{xu2013jstill}. Given the concise syntax characteristics of scripting languages like Python~\cite{ohm2022feasibility}, these methods demonstrate significant value in overcoming the limitations of traditional analysis techniques, providing new technical approaches for malicious code detection.

\noindent \textbf{Other Security Issues of PyPI Packages.}
In the domain of software package ecosystem security, PyPI faces diverse security threats. Typosquatting~\cite{taylor2020defending,liu2022exploring,kaplan2021survey} emerges as a primary attack vector, where attackers create malicious packages with names similar to popular ones. To counter these threats, the TypoGard~\cite{vu2020typosquatting} tool integrates lexical similarity models with package popularity metrics. Source code consistency verification~\cite{scalco2022feasibility} represents another key research direction, where py2src~\cite{vu2021py2src} establishes automatic mapping between PyPI packages and GitHub repositories, while LastPyMile~\cite{vu2021lastpymile} identifies potential malicious code injections through comparative analysis.
In cross-ecosystem detection research, Cerebro~\cite{zhang2025killing} builds a detection framework based on high-level behavior abstraction and BERT models, while Ruian~\cite{duan2020towards} achieves multi-ecosystem threat detection through registry metadata and system call analysis. Machine learning approaches demonstrate notable advantages, with Amalfi~\cite{sejfia2022practical} and PPD~\cite{liang2021malicious} enhancing malicious package detection through feature engineering and anomaly detection algorithms, respectively. Ea4mp~\cite{sun20241+} combines metadata and behavior to detect malicious PyPI packages~\cite{gao2025malguard}. However, existing methods rely on simple rules and black-box machine learning models, failing to effectively handle obfuscation and false positives that challenge practical deployment.
\section{Conclusion}
\label{sec:conclusion}

We present a knowledge-driven framework that transforms detection failures into behavioral knowledge for malicious package identification. By mining 304 discriminative patterns from 18,137 PyPI packages and integrating knowledge into our method, we achieve 99.50\% accuracy with only 2 false positives compared to 1,927-2,117 in existing tools. The framework maintains robust performance on obfuscated code with 98.28\% accuracy and demonstrates cross-ecosystem generalizability with 98.07\% accuracy on NPM packages. This semantic approach reduces false positives by 99\% while maintaining high performance against evolving threats.

\appendix

\section{Ethical Considerations}
\label{sec:ethics}

\wb{We conducted a stakeholder-based ethical analysis following the USENIX Security guidelines.}

\wb{\noindent\textbf{Stakeholder Identification.}
We identified the following stakeholders in our research: (1) \textit{Package registry maintainers} (PyPI, NPM, and mirror sites such as Tsinghua and Tencent), who host and distribute packages; (2) \textit{Developers and end users}, who may unknowingly install malicious packages; (3) \textit{Malicious package authors}, who create and distribute malicious packages; and (4) \textit{Security researchers}, who may build upon our work for future malware detection research.}

\wb{\noindent\textbf{Impact Analysis and Mitigations.}
For \textit{(1) package registry maintainers}, our data collection could potentially increase server load. To mitigate this, we implemented request throttling with 5-second intervals between consecutive downloads. Given PyPI's approximately 2.1 billion daily downloads~\cite{pypi_stats_2024}, our one-time collection of approximately 12,000 benign packages represents negligible load (<0.0006\% of daily traffic). Upon detecting potentially malicious packages during our real-world deployment, we followed PyPI's official security reporting guidelines. Each detected package underwent manual verification by our research team before submission to the PyPI security team via their official platform (\url{https://pypi.org/security/}). All 219 reported malicious packages encompassing 287 versions were confirmed and subsequently removed by PyPI maintainers, with 253 official acknowledgment letters received. We notified downstream mirror maintainers only after PyPI had confirmed and removed the malicious packages from the main repository.}

\wb{For \textit{(2) developers and end users}, our research aims to protect them by detecting malicious packages before they cause harm. No malicious package information was publicly disclosed through any channels prior to removal, preventing potential exploitation during the disclosure period.}

\wb{For \textit{(3) malicious package authors}, while our detection techniques could potentially inform evasion strategies, we believe the benefit of protecting the broader developer community outweighs this risk. \tool{} focuses on semantic-level detection rather than signature-based methods, making evasion more challenging.}

\wb{For \textit{(4) security researchers}, we release our dataset and tools to facilitate future research in malicious package detection.}

\wb{\noindent\textbf{Data Collection.}
All malicious packages in our study were obtained from publicly available datasets, including Guo et al.~\cite{guo2023empirical} and Backstabbers-Knife-Collection~\cite{ohm2020backstabber}, which are continuously maintained repositories of confirmed malicious packages. Benign packages were downloaded directly from PyPI.org and npmjs.com public registries. No personally identifiable information or sensitive user data was accessed or collected at any stage of this research. Additional details on dataset construction are provided in Section~\ref{subsec:study_dataset} and Section~\ref{subsec:eva_dataset}.}

\wb{\noindent\textbf{Publication Decision.}
We decided to publish this research because the benefits of enabling better malicious package detection outweigh the potential risks. Software supply chain attacks pose significant threats to developers and organizations worldwide, and our work provides practical tools and insights to defend against such attacks. The behavioral patterns we identified can help the security community develop more robust detection mechanisms.}

\section{Open Science}
\label{sec:openscience}

Our paper fully adheres to the open science policy introduced by USENIX Security. All code and experimental datasets are available at \url{https://doi.org/10.5281/zenodo.17929520}.

\section{Acknowledgments}
\label{sec:acknowledgment}

This research is supported by the National Research Foundation, Singapore, and DSO National Laboratories under the AI Singapore Programme (AISG Award No: AISG4-GC-2023-008-1B); by the National Research Foundation Singapore and the Cyber Security Agency under the National Cybersecurity R\&D Programme (NCRP25-P04-TAICeN); and by the Prime Minister’s Office, Singapore under the Campus for Research Excellence and Technological Enterprise (CREATE) Programme. Any opinions, findings and conclusions, or recommendations expressed in these materials are those of the author(s) and do not reflect the views of the National Research Foundation, Singapore, Cyber Security Agency of Singapore, Singapore.

\bibliographystyle{IEEEtran}
\bibliography{ref}

\end{document}